\RequirePackage{fix-cm}
\documentclass[twocolumn,epjc3]{svjour3}
\smartqed  
\usepackage{amsmath}
\usepackage{amssymb}
\usepackage{amssymb,bbold}
\usepackage{kantlipsum,widetext}

\RequirePackage{graphicx}
\usepackage{subfig}
\RequirePackage[numbers,sort&compress]{natbib}

\usepackage{lineno,hyperref}
\hypersetup{colorlinks=true,linkcolor=blue,citecolor=blue,filecolor=blue,urlcolor=blue,breaklinks=true}

\journalname{Eur. Phys. J. C}
\begin{document}

\title{Non-radial oscillations and global stellar properties of anisotropic compact stars using realistic equations of state
}


\author{Elvis J. Aquino Curi\thanksref{e4,addr1}\and Luis B. Castro\thanksref{e1,addr1} \and Cesar V. Flores\thanksref{e2,addr2,addr1} \and C\'{e}sar H. Lenzi\thanksref{e3,addr3} 
}

\thankstext{e4}{e-mail: joemaxell@gmail.com}
\thankstext{e1}{e-mail: lrb.castro@ufma.br, bcastro.luisr@gmail.com}
\thankstext{e2}{e-mail: cesarovfsky@gmail.com}
\thankstext{e3}{e-mail: chlenzi1980@gmail.com }


\institute{Departamento de F\'{\i}sica - CCET, Universidade Federal do Maranh\~{a}o (UFMA), Campus Universit\'{a}rio do Bacanga, CEP 65080-805, S\~{a}o Lu\'{\i}s, MA, Brazil.
\label{addr1}  \and Centro de Ci\^{e}ncias Exatas, Naturais e Tecnol\'{o}gicas - CCENT, Universidade Estadual da Regi\~{a}o Tocantina do Maranh\~{a}o (UEMASUL), CEP 65901-480, Imperatriz, MA, Brazil. 
\label{addr2} \and Departamento de F\'{\i}sica, Instituto Tecnol\'{o}gico de Aeron\'{a}utica, DCTA, CEP 12228-900, S\~{a}o Jos\'{e} dos Campos, SP, Brazil. \label{addr3}
}

\date{Received: date / Accepted: date}

\maketitle

\begin{abstract} In this work, we have made a systematic study of how the gravitational wave frequency of the fundamental mode from compact stars is affected by aniso\-tropic effects using realistic equations of state. Our study is an extension of the seminal research performed by Doneva [Phys. Rev. D 85 (2012) 124023], where a polytropic equation of state was used. To achieve our objective, we considered compact stars which were built by using equations of state in the framework of a relativistic mean field theory for the case of hadronic stars and in the framework of the MIT model for the case of quark stars. In order to obtain some pertinent information that could give us the possibility to detect the anisotropy in compact stars, we also studied and analized the behaviour of various global stellar quantities, e.g., gravitational redshift, stellar mass, radius, among others. We concluded that the anisotropic effects can have important consequences, which are strongly related to the aniso\-tropic parameter and the equation of state of high density matter. Additionally, a comparison with observational data has been made and we have shown that the anisotropic parameter $\lambda$ can be used as a tuning parameter to reproduce mass and radius observational data of neutron stars.

\end{abstract}

\section{Introduction}
\label{intro}

It is well know that compact stars are astrophysical objects in whose interior matter can be found in extreme conditions, i.e., high densities, intense gravitational fields, strong magnetic fields and fast rotation. The high density matter in neutron stars can be studied using an equation of state (EoS), which is used as the main ingredient for the numerical integration of the stellar structure equations, the so called Tolman-Oppenheimer-Volkoff (TOV) equations. By using this system of equations, important quantities can be theoretically obtained, e.g., the mass and radius of a compact star, compacticity, gravitational redshift, among other. Recently there have been efforts in order to obtain better observational constraints on the mass an radius of compact stars and as consequence we have seen an improvement on the knowledge of the high density matter inside compact stars. 

Another important source of information about compact stars comes from their gravitational wave emission. This information can be observed from two main channels: binary systems, in which a neutron star is one of the binary components and from neutron star oscillations. In the second case the neutron star is  considered as an isolated system and can be set in an oscillatory state as a consequence of some internal or external perturbation. Those oscillation modes constitute the fingerprint of the compact star, each mode with a natural frequency corresponding to some characteristic or dynamics of the fluid. Among all possible oscillatory frequencies we have the fundamental mode which has been focus of intense investigation because it can be detected by future third generation gravitational wave detectors or because its effects can be observed when there exist resonance in a binary system.

All previously mentioned characteristics about compact stars are frequently studied for the case of a perfect isotropic fluid, but, at the same time, it is also very well know that the knowledge of the EoS in the inner core of neutron stars is very elusive. That difficulty has its origins in the uncertainty of the nuclear EoS at extreme high densities, and for this reason some authors have proposed  that inside neutron stars could exist deconfined quark matter and others authors have even proposed the existence of anisotropic effects. The existence of anisotropy can be justified by many reasons, among them we have the possibility of a solid core or the presence of strong magnetic fields, and by this reason \cite{1997PhR...286...53H} the possibility of anisotropy inside neutron stars has been object of recent study, e.g., \cite{2000PhR...328..237H,1972PhRvL..29..823S,PRD85:124023:2012}. Recently, the conditions for the (in)stability of the isotropic pressure condition in collapsing spherically symmetric, dissipative fluid distributions have been investigated in \cite{PRD101:104024:2020}. In this seminal work, the author concluded that dissipative fluxes, and/or energy density inhomogeneities and/or the appearance of shear in the fluid flow, force any initially isotropic configuration to abandon such a condition, generating anisotropy in the pressure, i.e an initial fluid configuration with isotropic pressure would tend to develop an anisotropic pressure as it evolves, under conditions expected in stellar evolution. The\-refore, we are forced to consider pressure anisotropy whenever relativistic fluids are involved. 

It is very well know that the use of polytropic EoS reduce numerical computations and describe consistently many global properties of compact objects \cite{Tooper1964}, therefore polytropic EoS can serve as an extension to produce realistic EoS \citep{Yunes2014}. For example, we can use a polytropic index of $N = 0.5\,, 0.6$ in order to obtain a maximum neutron star mass above two solar masses and radius in the range of the NICER observations \cite{NICER}. Additionally, several polytropic equations can be used to model different densities inside neutron stars, this is called the piecewise polytropic approximation. In Ref.~\cite{PRD85:124023:2012}, the authors used a polytropic EoS to study the non-radial oscillations (in Cowling approximation) of neutron stars in the presence of anisotropic pressure. Our objective here is to complement that work. To achieve this goal, we use EoS in the framework of a relativistic mean field theory \cite{WALECKA1986} for hadronic stars and in the framework of the Mit bag model \cite{PRD9:3471:1974,PRD30:2379:1984} for quark stars.

The paper has the following structure: In section \ref{GTOV} we recall the equilibrium configuration of anisotropic compact stars. In section \ref{COWLING} we make a short explanation of the equations governing the oscillations of anisotropic stars. In section \ref{EOS} we describe realistic equations of state that are used in order to model compact stars. In section \ref{RESU} we discuss our results and perform a comparison of our results with observational data. Finally in section \ref{CONCLU} we give our final conclusions. Through our discussion we use relativistic units $c=1$ and $G=1$, where $c$ is the speed of light and $G$ is the gravitational constant, respectively.

\section{Generalized Tolman-Oppenheimer-Volkoff (TOV) equations for an anisotropic fluid distribution} 
\label{GTOV}

The Tolman-Oppenheimer-Volkoff (TOV) equations for an static and spherical star are derived from the standard  general relativity Einstein field equations~\cite{PR55:364:1939,PR55:374:1939}
\begin{equation}
G_{\mu \nu}=R_{\mu \nu}-\frac{1}{2}Rg_{\mu \nu}=8\pi T_{\mu \nu},
\end{equation}
\noindent where $R_{\mu \nu}$ is the Ricci tensor, $R=g^{\mu \nu}R_{\mu \nu}$ is the scalar curvature, $T_{\mu \nu}$ is the energy-momentum tensor and $g_{\mu \nu}$ is the metric given by the coefficients of the line element below
\begin{equation}\label{difel}
	ds^2=-e^{2\Phi(r)}dt^2+e^{2\Lambda(r)}dr^2+r^2(d\theta^2+\sin^2\theta d\phi^2).
\end{equation}
\noindent The anisotropic fluid distribution in spherical symmetry can be represented by the following energy-momen\-tum tensor
\begin{equation}
T_{\mu\nu}=\rho u_{\mu}u_{\nu}+p k_{\mu}k_{\nu}+q\left( g_{\mu\nu}+u_{\mu}u_{\nu}-k_{\mu}k_{\nu} \right),\
\end{equation}
\noindent where $u_{\mu}$ is the fluid $4$-velocity ($u^{\mu}u_{\mu}=-1$), $\rho$ is the fluid energy density, $k^{\mu}$ is the unit radial vector ($k^{\mu}k_{\mu}=1$) and  $u^{\mu}k_{\mu}=0$. The quantities $p$ and $q$ are the radial and tangential pressures, respectively. 

By solving the Einstein field equations for the aniso\-tropic fluid distribution we can obtain the differential equations that govern the equilibrium configuration of the star
\begin{eqnarray}
\frac{dm}{dr}&=&4\pi r^2\rho,\\
\frac{dp}{dr}&=&-(\rho+p)\frac{d\Phi}{dr}-\frac{2\sigma}{r},\\
\frac{d\Phi}{dr}&=&\frac{m+4\pi r^3p}{r(r-2m)},\\
\nonumber
\end{eqnarray}
\noindent where $m(r)$, $p(r)$, $\rho(r)$, $\Phi (r)$, $\sigma(r)$ are quantities that depend on the radial coordinate $r$ and $\sigma=p-q$ is the anisotropic profile. For our purposes this profile will be defined as \cite{PRD85:124023:2012}
\begin{equation}
\sigma = \lambda p \mu
=\lambda p\left(1-\mathrm{e}^{-2\Lambda}\right)\,, 
\end{equation}
\noindent where $\lambda$ is a dimensionless parameter which controls the anisotropy and $\mu$ is a quasilocal variable. 

As we can see, there are 3 equations and 4 unknown functions. For this reason, in order to solve this system of equations, we need an equation of state $p(\rho)$ and additionally suitable boundary conditions.

The integration of the differential equations above is made with the appropriate boundary conditions at $r=0$:
\begin{equation}
\begin{split}
& m(0)=0,\quad p(0)=p_c,\quad \Phi(0)=\Phi_c, 
\end{split}
\end{equation}
\noindent where $p_c$ is the central pressure and $\Phi_c$ is the metric field value at the star center. The integration begins with the initial conditions mentioned above and proceeds with tiny steps in the radius $r$. Then when the integrated pressure is zero, the numerical integration stops and the radius $R$ and mass $M$ of the star are obtained.

It is also important to comment that the third boundary condition is a guess for the metric field at the center, therefore the metric has to be renormalized using the value of the vacuum Schwarzschild metric at $r=R$. That procedure is necessary, because we need the correct metric values in order to obtain the coefficients of the oscillations equations, which are going to be discussed in the next section.

\section{Cowling approximation for non-radial perturbations}
\label{COWLING}

This approximation was introduced by Cowling \cite{MNRAS101:367:1941} for Newtonian stars and by McDermott for the case of neutron stars. This approximation consists in ignoring the coupling between the fluid motions and the metric perturbations. It means that we can completely neglect the metric perturbations (spacetime metric is kept fixed). In the next lines we will briefly explain how the equations that describe the perturbations equations in the Cowling formalism are obtained. For further details see \cite{PRD85:124023:2012},

To obtain the oscillation equations, we have to consider the perturbation of the conservation equation for the energy momentum tensor
\begin{equation}
\nabla_{\nu}\delta T^{\nu}_{\mu}=0.
\label{perT}
\end{equation}
\noindent We can project Eq. (\ref{perT}) along the $u^{\mu}$ and obtain
\begin{eqnarray}\label{PPU}
u^{\nu}\nabla_{\nu}\delta\rho + \nabla_{\nu}\left\{\left[(\rho +
q)\delta^{\nu}_{\mu}+ \sigma k^{\nu}k_{\mu}\right]\delta
u^{\mu}\right\} \nonumber \\ 
 +\left(\rho + q\right)a_{\nu}\delta u^{\nu} +
\nabla_{\nu}u_{\mu}\delta \,\left(\sigma k^{\nu}k^{\mu}\right)=0.
\end{eqnarray}
\noindent Projecting orthogonally to the background $4$-velocity by using the operator ${\cal P}_{\mu}^{\nu}=\delta^{\nu}_{\mu} + u^{\nu}u_{\mu}$, we obtain
\begin{eqnarray}\label{POE}
\left(\delta\rho + \delta q\right)a_{\mu} + \left(\rho +
q\right)u^{\nu} \left(\nabla_{\nu}\delta u_{\mu} -
\nabla_{\mu}\delta u_{\nu} \right) 
\nonumber \\ 
+\nabla_{\mu}\delta q + u_{\mu}
u^{\nu}\nabla_{\nu}\delta q + {\cal
	P}_{\mu}^{\nu}\nabla_{\alpha}\delta\left(\sigma k^{\alpha }k_{\nu}
\right) =0,
\end{eqnarray}
\noindent where $a_{\mu}=u^{\nu}\nabla_{\nu}u_{\mu}$ is the background 4-acceleration.

As the star is oscillating, the elements of fluid move from their equilibrium position, that motion is represented by the displacements $\xi^{i}$, where $ i=1,2,3=r,\theta, \phi$. Additionally an integrability condition can be obtained from Eq. (\ref{POE}) from which the correct expressions for $\xi_{\theta}$ and $\xi_{\phi}$ can be calculated. 

Using the Eq. (\ref{PPU}) and $\xi^{i}$, we can explicitly write down the expressions for the density and pressure perturbations
\begin{equation}
\begin{split}
\delta\rho=&-\left( \rho+p \right)\left[ \mathrm{e}^{-\Lambda}\frac{W^{\prime}}{r^2}+\frac{l(l+1)}{r^2}V \right]Y_{lm}\\
 & -\frac{d\rho}{dr}\mathrm{e}^{-\Lambda}\frac{W}{r^2}Y_{lm}+\frac{2\sigma}{r^3}\mathrm{e}^{-\Lambda}WY_{lm}\\
 & +\sigma\frac{l(l+1)}{r^2}VY_{lm}\,,
\end{split}
\end{equation}
\begin{equation}
\begin{split}
\delta p=&-\frac{dp}{d\rho}\left\{ (\rho+p)\left[ \mathrm{e}^{-\Lambda}\frac{W^{\prime}}{r^2}+\frac{l(l+1)}{r^2}V \right]
-\frac{2\sigma}{r^3}\mathrm{e}^{-\Lambda}W\right. \\
 & -\left.\sigma\frac{l(l+1)}{r^2}V\right\}Y_{lm}-\frac{dp}{dr}\mathrm{e}^{-\Lambda}\frac{W}{r^2}Y_{lm}, \
\end{split}
\end{equation}
\noindent where $Y_{lm}$ are the spherical harmonics, $W$ and $V$ are perturbation functions and the symbol prime denotes the derivative with respect to the radial coordinate $r$. 

For the perturbation of the anisotropic pressure $\sigma=\sigma(p,\mu)$, we have
\begin{equation}
\delta\sigma=\frac{\partial\sigma}{\partial p}\delta p\,,
\end{equation}
\noindent where was considered $\delta\mu=0$.

After considering a harmonic dependence on time for the perturbation functions $W(r,t)=W(r)\mathrm{e}^{i\omega t}$ and $V(r,t)=V(r)\mathrm{e}^{i\omega t}$, where $\omega$ is the oscillation frequency, the oscillation equations in the Cowling approximation can be obtained 
\begin{equation}
\begin{split}
W^{\prime}=&\frac{d\rho}{dp}\left[ \omega^2\frac{\rho+p-\sigma}{\rho+p}\left( 1-\frac{\partial\sigma}{\partial p} \right)^{-1}\mathrm{e}^{\Lambda-2\Phi}r^{2}V+\Phi^{\prime}W \right]\\
 &-l(l+1)\mathrm{e}^{\Lambda}V+\frac{\sigma}{\rho+p}\left[ \frac{2}{r}\left( 1+\frac{d\rho}{dp} \right)W \right.\\
 &+\left.l(l+1)\mathrm{e}^{\Lambda}V\right]\,,
\end{split}
\end{equation}
\begin{equation}
\begin{split}
V^{\prime}=&2V\Phi^{\prime}-\left(1-\frac{\partial\sigma}{\partial p}\right)\frac{\rho+p}{\rho+p-\sigma}\frac{\mathrm{e}^{\Lambda}}{r^2}W\\
&+\left[ \frac{\sigma^{\prime}}{\rho+p-\sigma}+\left(\frac{d\rho}{dp}+1\right)\frac{\sigma}{\rho+p-\sigma}
\left( \Phi^{\prime}+\frac{2}{r} \right) \right.\\
&\left.-\frac{2}{r}\frac{\partial\sigma}{\partial p}-\left(1-\frac{\partial\sigma}{\partial p}\right)^{-1}
\left( \frac{\partial^{2}\sigma}{\partial p^2}p^{\prime}+\frac{\partial^{2}\sigma}{\partial p\partial\mu}\mu^{\prime} \right)\right] V.
\end{split}
\end{equation}

In order to solve the equations above we have to consider boundary conditions at the center and surface of the star. The boundary condition at the star surface furnishes
\begin{equation}
\begin{split}
\omega^{2}\frac{\rho+p-\sigma}{\rho+p}\left( 1-\frac{\partial\sigma}{\partial p} \right)^{-1}\mathrm{e}^{-2\Phi}V\\
+\left( \Phi^{\prime}+\frac{2}{r}\frac{\sigma}{\rho+p} \right)\mathrm{e}^{-\Lambda}\frac{W}{r^2}=0\,,
\end{split}
\end{equation}
\noindent and the boundary condition at the star center ($r=0$) satisfies
\begin{equation}
\tilde{W}=-l\tilde{V}\,,
\end{equation}
\noindent where it was introduced the new functions defined by $W=\tilde{W}r^{l+1}$ and $V=\tilde{V}r^{l}$. Hereafter, for all our results, we will consider $l=2$, i.e., we restrict to the quadrupolar modes.

\section{Equation of state}
\label{EOS}

As mentioned in the Introduction, depending on their possible interior composition, neutron stars can be classified as hadronic stars with or without hyperons, hybrid stars containing hadronic and quark phases and quark stars (strange star). In this work, we will consider hadronic stars without hyperons and quark stars.    

\subsection{Hadronic stars}

For the description of the EoS of hadronic matter, we employ the relativistic nonlinear Walecka model (NLW) \cite{WALECKA1986}. The total Lagrangian density reads
\begin{equation}\label{lagtotal}
\mathcal{L}_{H}=\sum_{b}\mathcal{L}_{b}+\mathcal{L}_{m}+\sum_{L}\mathcal{L}_{L}\,,
\end{equation}
\noindent where $\mathcal{L}_{b}$, $\mathcal{L}_{m}$ and $\mathcal{L}_{L}$ are the baryons, mesons and leptons Lagrangians, respectively, and are given by
\begin{eqnarray}\label{lb}
    \mathcal{L}_{b} &=& \bar{\psi}_{b}\left(i\gamma_{\mu}\partial^{\mu}-m_{b}+g_{\sigma b}\sigma \right. \nonumber \\
    && \left. -g_{\omega b}\gamma_{\mu}\omega^{\mu}-\frac{1}{2}g_{\rho b}\gamma_{\mu}\vec{\tau}\cdot\vec{\rho}^{\mu}\right)\psi_{b}\,,
\end{eqnarray}
\begin{eqnarray}\label{lm}
   \mathcal{L}_{m} &=& \frac{1}{2}\left(\partial_{\mu}\sigma\partial^{\mu}\sigma-m_{\sigma}^{2}\sigma^{2}\right)-U(\sigma)+
    \frac{1}{2}m_{\omega}^{2}\omega_{\mu}\omega^{\mu} \nonumber \\
   && -\frac{1}{4}\omega_{\mu\nu}\omega^{\mu\nu}+
    \frac{1}{2}m_{\rho}^{2}\vec{\rho}_{\mu}\cdot\vec{\rho}^{\mu}-\frac{1}{4}\vec{\rho}^{\mu\nu}\cdot\vec{\rho}_{\mu\nu} \,,
\end{eqnarray}
\begin{equation}\label{ll}
    \mathcal{L}_{L}=\bar{\psi}_{L}\left(i\gamma_{\mu}\partial^{\mu}-m_{L}\right)\psi_{L} \,,
\end{equation}
\noindent where the $b$ sum runs over the nucleons $b\equiv p,n$, and $\psi_{b}$ is the corresponding baryon Dirac field, whose interactions are mediated by the $\sigma$ scalar, $\omega_{\mu}$ isoscalar-vector, and $\rho_{\mu}$ isovector-vector meson fields. The baryon mass and isospin are denoted by $m_{b}$ and $\vec{\tau}$, respectively. The term
\begin{equation}
U(\sigma)=\frac{1}{3}\,bm_{n}(g_{\sigma}\sigma)^{3}-\frac{1}{4}\,c(g_{\sigma}\sigma)^{4}\,,
\end{equation}
\noindent denotes the scalar self-interactions. The mesonic tensors are given by their usual expressions $\omega_{\mu\nu}=\partial_{\mu}\omega_{\nu}-\partial_{\nu}\omega_{\mu}$ and $\vec{\rho}_{\mu\nu}=\partial_{\mu}\vec{\rho}_{\nu}-\partial_{\nu}\vec{\rho}_{\mu}-g_{\rho b}\left( \vec{\rho}_{\mu}\times\vec{\rho}_{\nu} \right)$. The $L$ sum runs over the two lightest leptons $L\equiv e,\mu$ and $\psi_{L}$ is the lepton Dirac field. For more details about the EoS obtained from the Lagrangian (\ref{lagtotal}) see Ref. \cite{JPG40:035201:2013} and references therein. 

The coupling constants adopted in this work are given by the GM1 \cite{PRL67:2414:1991} and NL3 \cite{PRC55:540:1997} parametrizations (see Table \ref{table:1}).

\begin{table}[h]
 \centering
\begin{tabular}{c c c} \hline
 Set & GM1 & NL3\\  \hline
$m_{\sigma}$ (MeV)& 512 & 508.194 \\  
$m_{\omega}$ (MeV)& 783 & 782.501 \\  
$m_{\rho}$ (MeV)& 770 & 763  \\  \hline
$g_{\sigma}$ & 8.91 & 10.217 \\  
$g_{\omega}$ & 10.61 & 12.868 \\
$g_{\rho}$ & 8.196 & 8.948 \\
b & 0.002947 & 0.002055 \\
c & -0.001070 & -0.002651 \\ \hline
\end{tabular}
 \caption{\label{table:1} Coupling constants for the GM1 and NL3 parametrizations.}
\end{table}

\subsection{Quark stars}

For the description of the EoS of quark matter, we consider the phenomelogical MIT bag model, which characterizes a degenerated Fermi gas of quarks up, down and strange \cite{PRD9:3471:1974,PRD30:2379:1984}. The EoS is given by a linear relationship which relates the pressure and the energy density, that is 
\begin{equation}
p=\frac{1}{3}\left( \rho-4B \right)\,,
\end{equation}
\noindent where $B$ is the bag constant. For stable strange quark matter the bag constant values ranges from $57$\, $\mathrm{MeV}\mathrm{fm}^{-3}$ to $92$\, $\mathrm{MeV}\mathrm{fm}^{-3}$ \cite{SCHMITT2010}. It is worth to mention that a more recently work reports a slightly different range $58.926$\, $\mathrm{MeV}\mathrm{fm}^{-3}$ to $91.5$\, $\mathrm{MeV}\mathrm{fm}^{-3}$ \cite{LRR20:7:2017}.

\section{Results}
\label{RESU}

We used the non-radial oscillations equations from Doneva's work \cite{PRD85:124023:2012}. At first, before using realistic equations of state, we reproduced Doneva's results and those calculations can be considered as a strong test for our code. In this section we show different profiles: mass-radius, frequency-mass, mass-redshift, mass-density, normalized frequency-mass, and frequency-mean density of compact stars. Those results for hadronic stars were obtained using two parametrizations: GM1 and NL3 (see table \ref{table:1}) and for quark stars were obtained using two different values for the bag constant: $B=60$\, $\mathrm{MeV}\mathrm{fm}^{-3}$ and $B=90$\, $\mathrm{MeV}\mathrm{fm}^{-3}$. For reasons of comparison we also show the case of isotropic stars, which correspond to the black full lines ($\lambda=0$). 

\subsection{Hadronic stars}

\begin{figure*}[]
\subfloat[]{\label{fig:h1a}\includegraphics[angle=-90,width=0.47\textwidth]{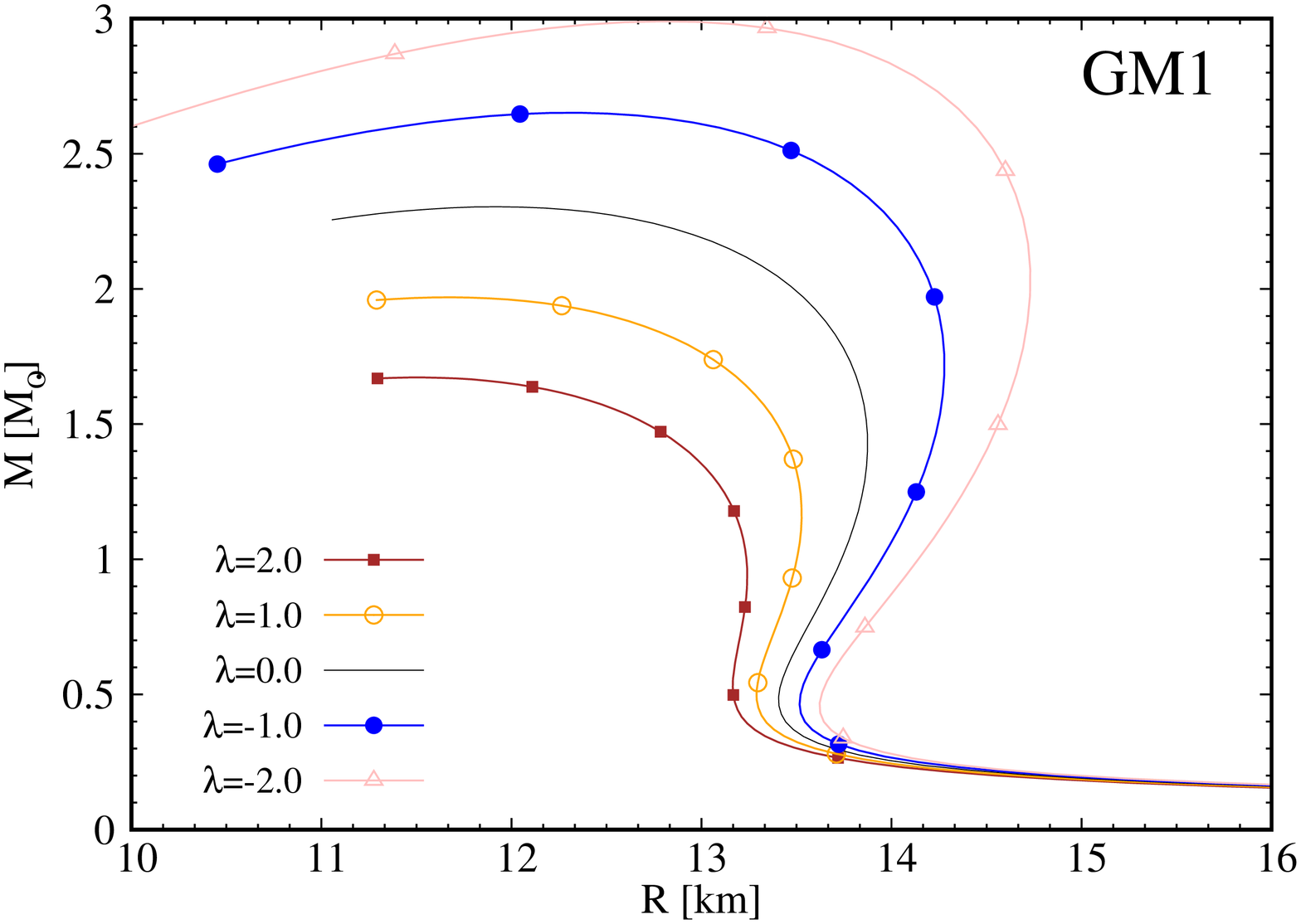}}\hspace{\fill}
\subfloat[]{\label{fig:h1b}\includegraphics[angle=-90,width=0.47\textwidth]{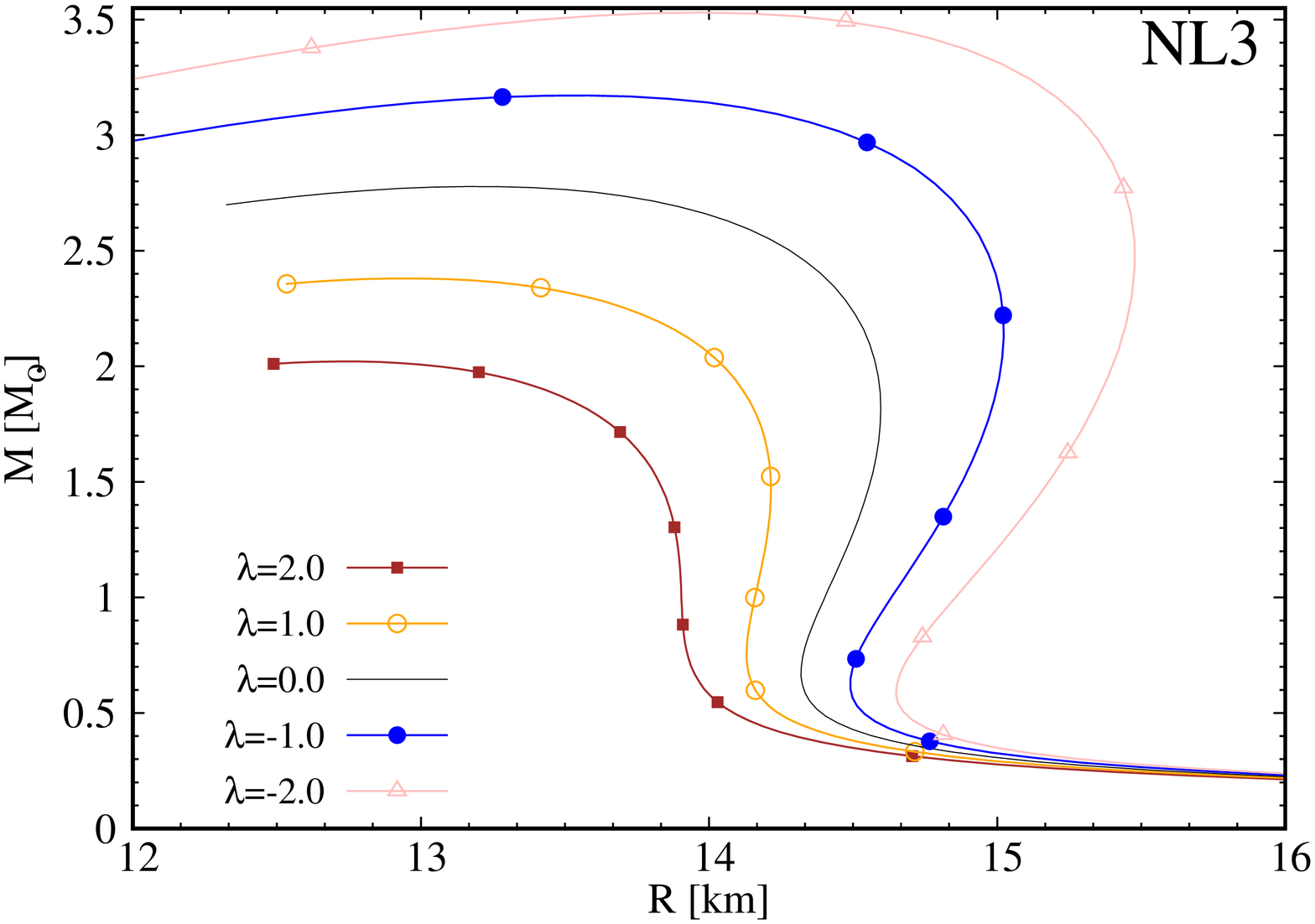}}
\caption{Mass-radius curves for anisotropic hadronic stars using the GM1 (left panel) and NL3 (right panel) parametrizations for different values of the parameter $\lambda$. Black lines ($\lambda = 0$) correspond to isotropic hadronic stars.}
\label{fig:h1}
\end{figure*}

As a first result, in Fig. \ref{fig:h1} we observe, as expected, that the maximum mass of the models are different for each value of the $\lambda$ parameter. We remark that smaller negative $\lambda$ values favor the pressure of the fluid to support stars with larger masses and in contrast, a large positive $\lambda$ value produces stellar configurations with smaller masses. When the GM1 model is employed, we observe that for hadronic masses below $0.3~M_{\odot}$ and radii greater than $13.2$~km, the anisotropy does not have important effects. A similar situation can be seen for the NL3 model, this behavior can be observed for stars with masses smaller than $0.3~M_{\odot}$ and radii greater than $14.5$~km. In conclusion, hadronic stars with smaller masses are not affected by anisotropy when they are described by the GM1 or the NL3 model. Therefore, if we want to detect anisotropic effects, we have to focus our attention on region of massive stars.

\begin{figure*}[]
\subfloat[]{\label{fig:h2a}\includegraphics[angle=-90,width=0.47\textwidth]{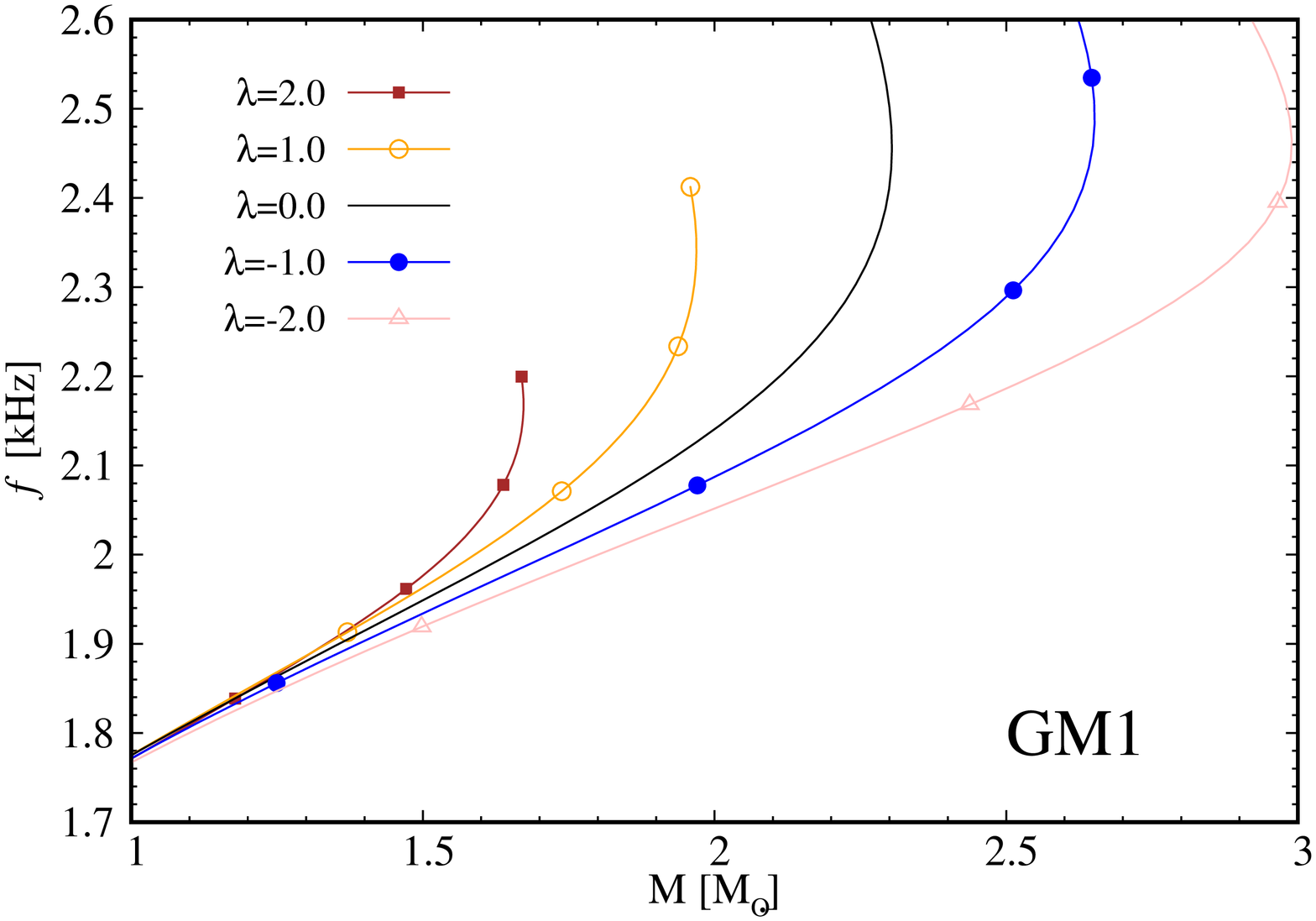}}\hspace{\fill}
\subfloat[]{\label{fig:h2b}\includegraphics[angle=-90,width=0.47\textwidth]{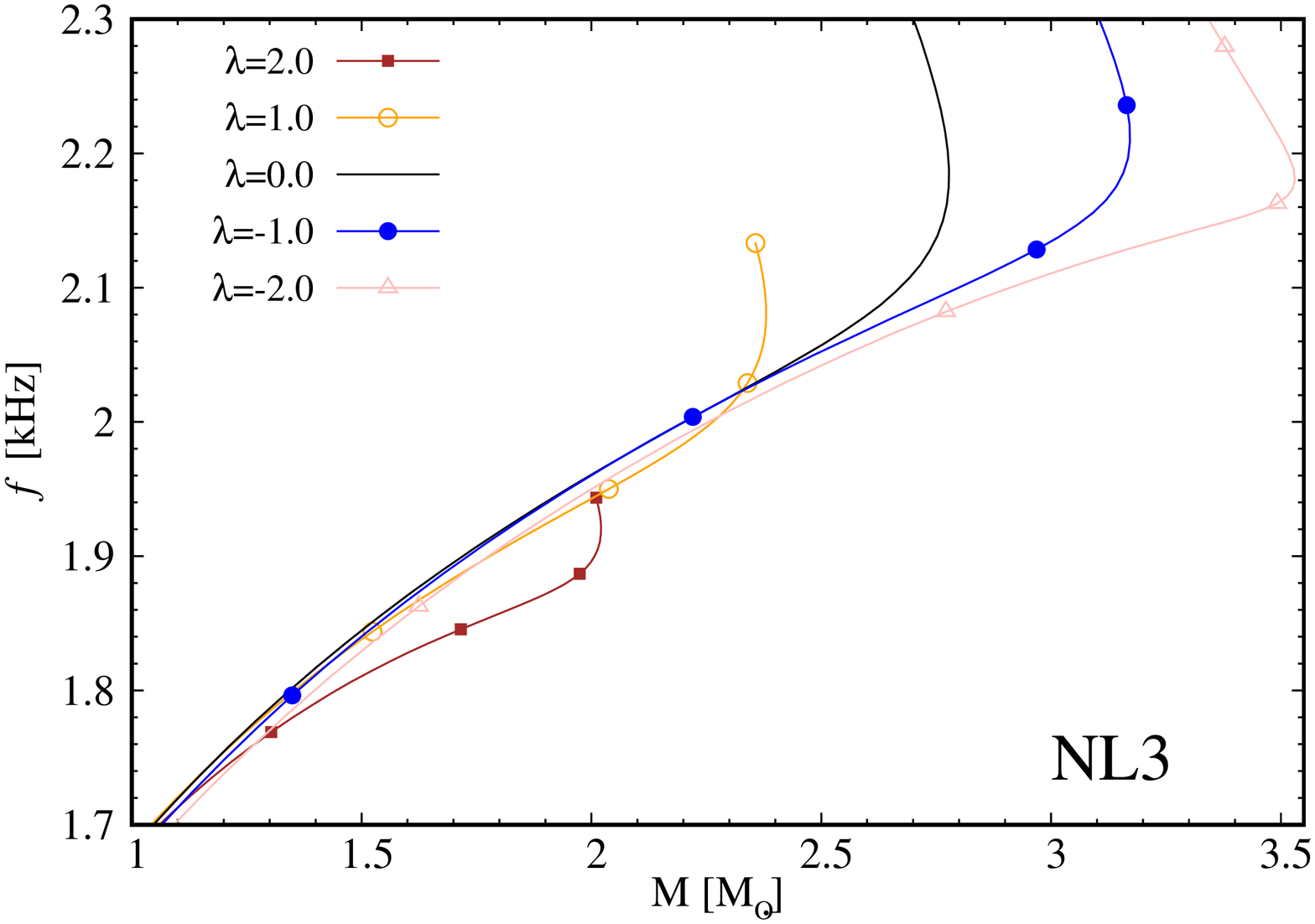}}
\caption{The frequency $f$ of the fundamental mode as a function of the mass $M$ using  GM1 (left panel) and NL3 (right panel) parametrizations. The results are shown for different values of the parameter $\lambda$.}
\label{fig:h2}
\end{figure*}

In the sequence, we show the results for the frequency of the fundamental mode as a function of the mass $M$, these results are presented on the Fig. \ref{fig:h2}. According to the GM1 parameterization (\ref{fig:h2a}), we observe that for masses below $1.5~M_{\odot}$, the fundamental mode frequencies of anisotropic hadronic stars do not have big difference when compared with  the frequencies of isotropic hadronic stars ($\lambda=0$). For masses above $1.5~M_{\odot}$, it is clear that as the $\vert\lambda\vert$ increases, the frequencies change significantly. The same behavior can be observed for hadronic stars build with the NL3 parametrization, as can be seen in the figure (\ref{fig:h2b}). We can conclude the GM1 model gives greater frequencies when compared with the NL3 model. 

\begin{figure*}[]
\subfloat[]{\label{fig:h3a}\includegraphics[angle=-90,width=0.47\textwidth]{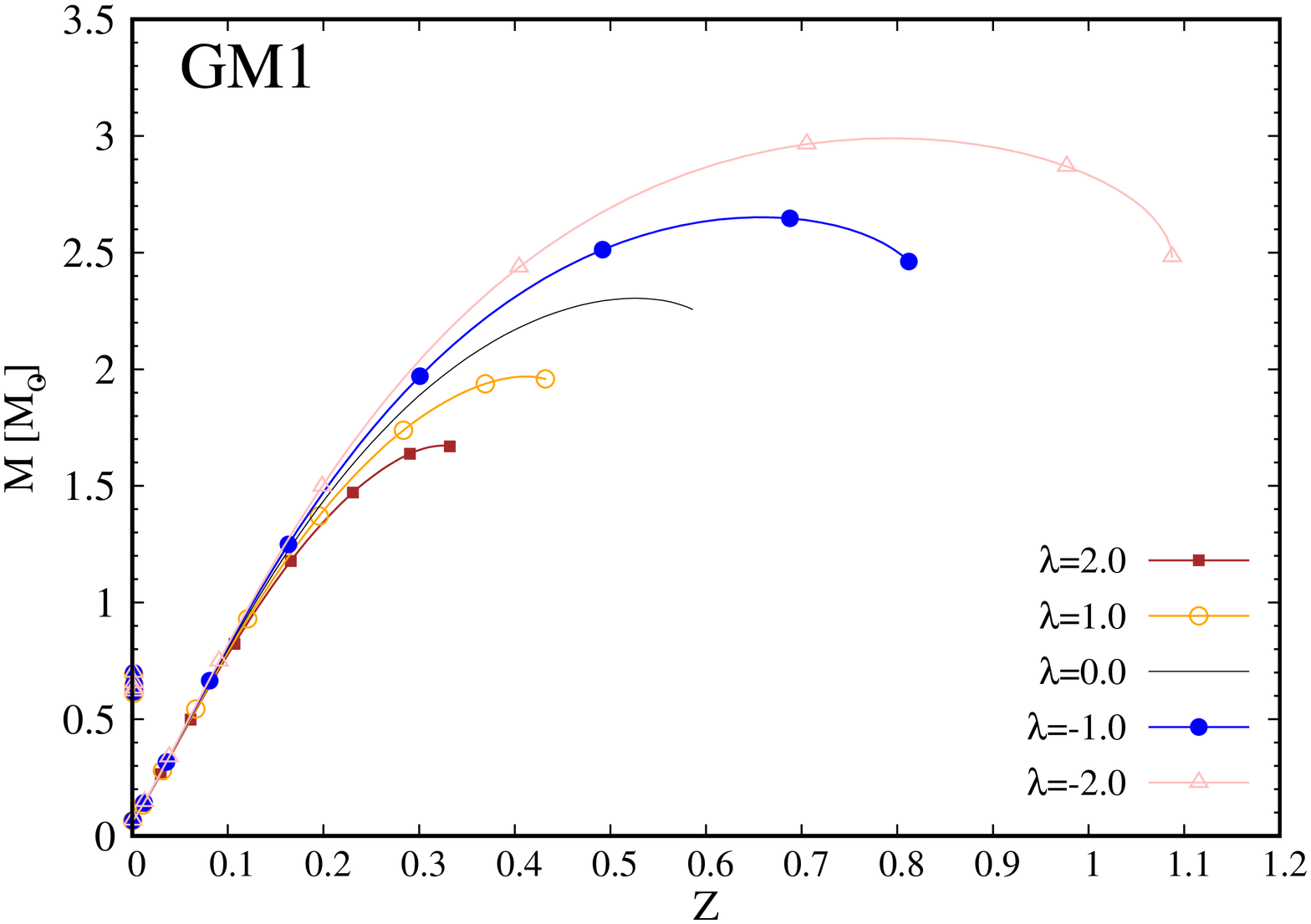}}\hspace{\fill}
\subfloat[]{\label{fig:h3b}\includegraphics[angle=-90,width=0.47\textwidth]{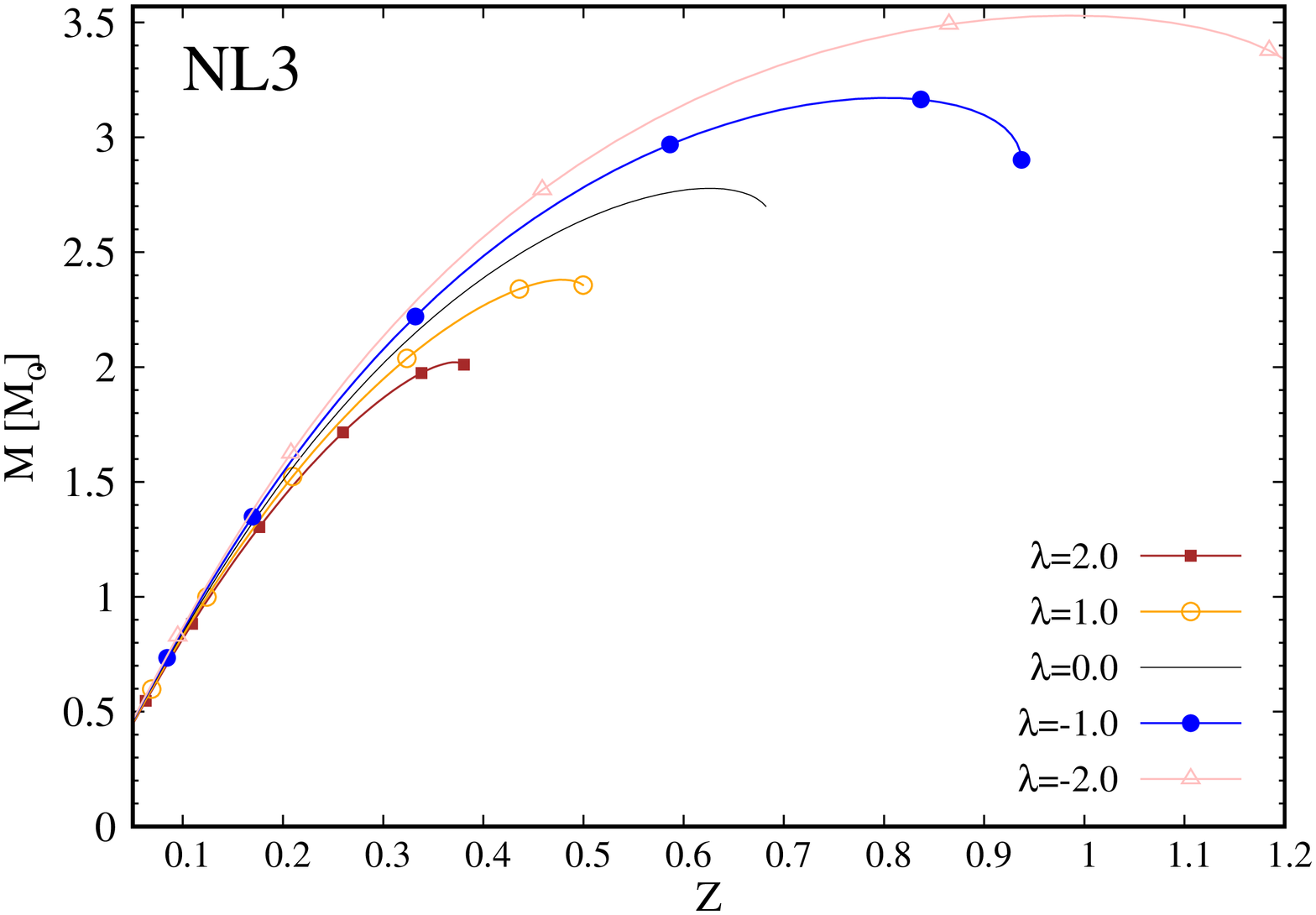}}
\caption{The total mass of the anisotropic hadronic stars as a function of the gravitational redshift $Z(R)=\mathrm{e}^{-\Phi(R)/2}-1$  for several values of the parameter $\lambda$. We use GM1 (left panel) and NL3 (right panel) parametrizations.}
\label{fig:h3}
\end{figure*}

As it is well know in general relativity, an important quantity is the surface gravitational redshift. In fact the gravitational redshift $Z$, which is predicted by Einstein's general relativity, expresses that light emitted from the surface of a compact object is expected to be displaced towards longer wavelengths of the electromagnetic spectrum.  The results for the gravitational redshift as a function of the mass are showed in figure \ref{fig:h3}, for different values of $\lambda$. From the two figures, we observe that the gravitational redshift increases linearly for masses smaller than $\approx 1.5~M_{\odot}$, this means that for low mass configurations, the values of the gravitational redshift becomes independent of the parameter $\lambda$. For stars with masses greater than $1.5~M_{\odot}$, the anisotropy produces significant changes for any $\lambda$. It is also clear that lower negative values of $\lambda$ give greater gravitational redshifts.

\begin{figure*}[]
\subfloat[]{\label{fig:h4a}\includegraphics[angle=-90,width=0.47\textwidth]{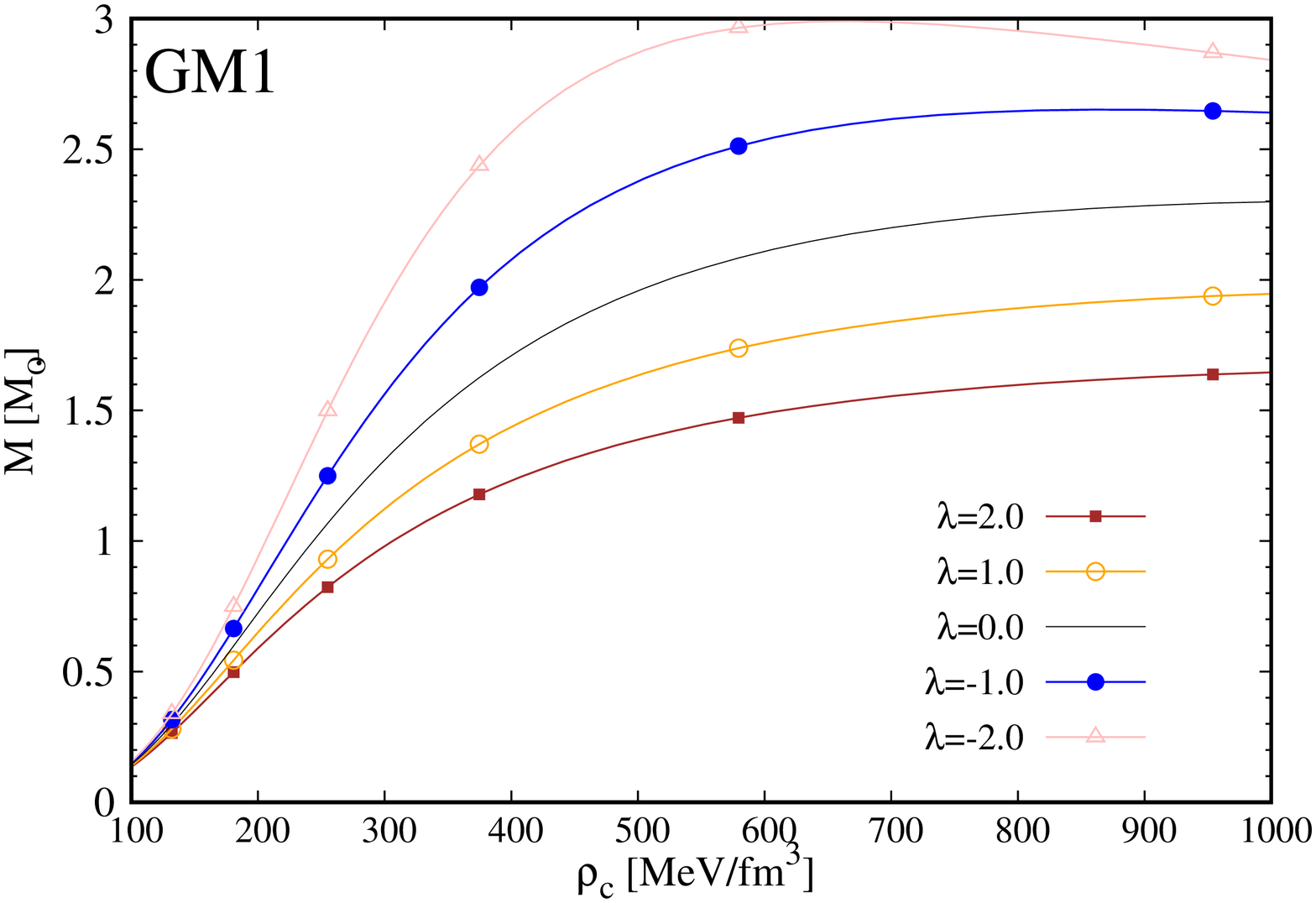}}\hspace{\fill}
\subfloat[]{\label{fig:h4b}\includegraphics[angle=-90,width=0.47\textwidth]{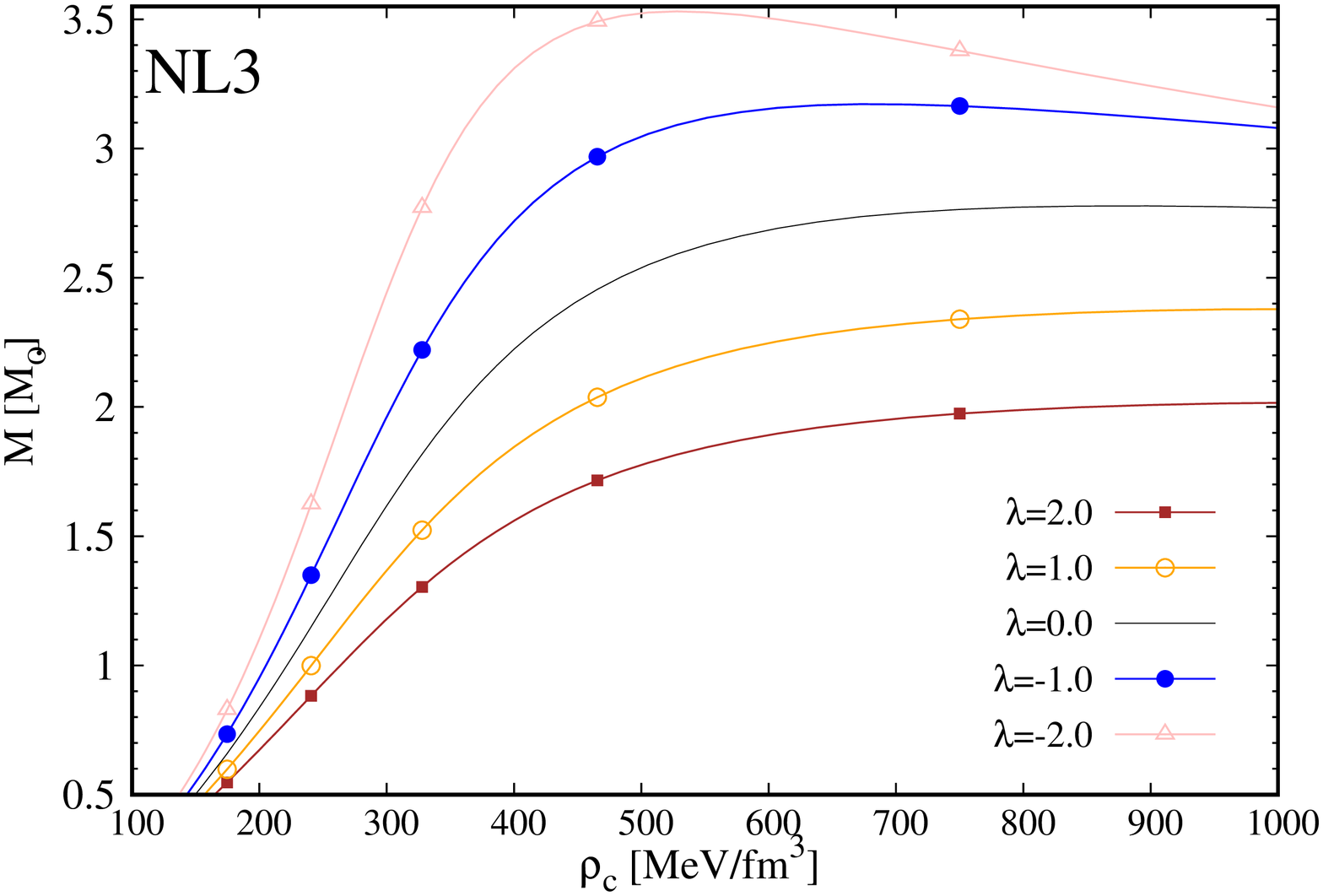}}
\caption{The total mass of anisotropic hadronic stars versus the central density $\rho_{c}$ using GM1 (left panel) and NL3 (right panel) parametrizations for different values of the parameter $\lambda$.}
\label{fig:h4}
\end{figure*}

In Fig. \ref{fig:h4}, the mass $M$ of anisotropic hadronic stars is shown as a function of the central density $\rho_c$. We observe that for both the GM1 and NL3 models, the properties of hadronic stars vary significantly as we change the parameter $\lambda$.  In fact, for central densities less than $200$\, $\mathrm{MeV}\mathrm{fm}^{-3}$, the masses of the anisotropic stars do not have a visible change, in contrast for values greater than $200$\, $\mathrm{MeV}\mathrm{fm}^{-3}$, the stellar masses change considerably. 

\begin{figure*}[]
\subfloat[]{\label{fig:h5a}\includegraphics[angle=-90,width=0.47\textwidth]{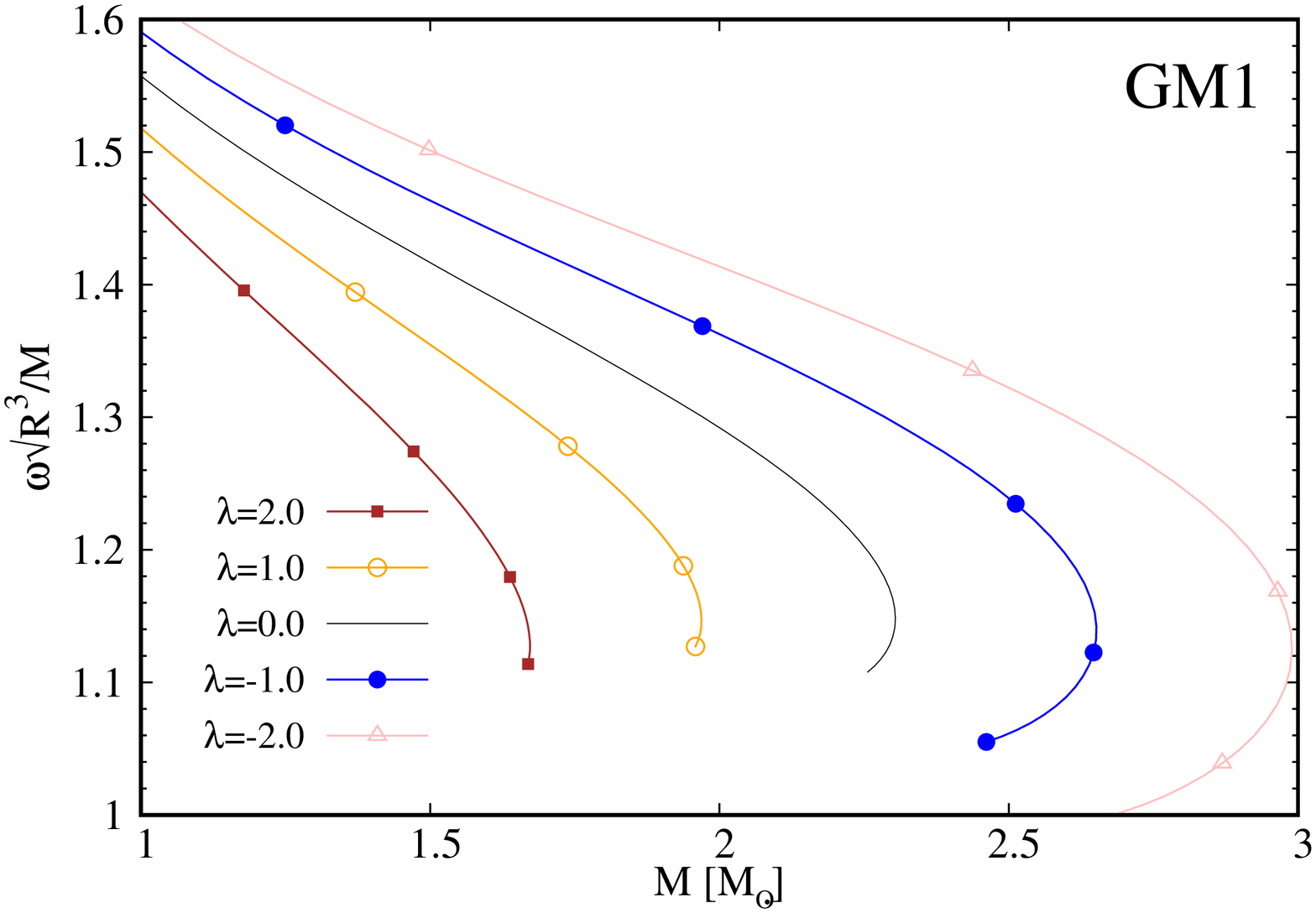}}\hspace{\fill}
\subfloat[]{\label{fig:h5b}\includegraphics[angle=-90,width=0.47\textwidth]{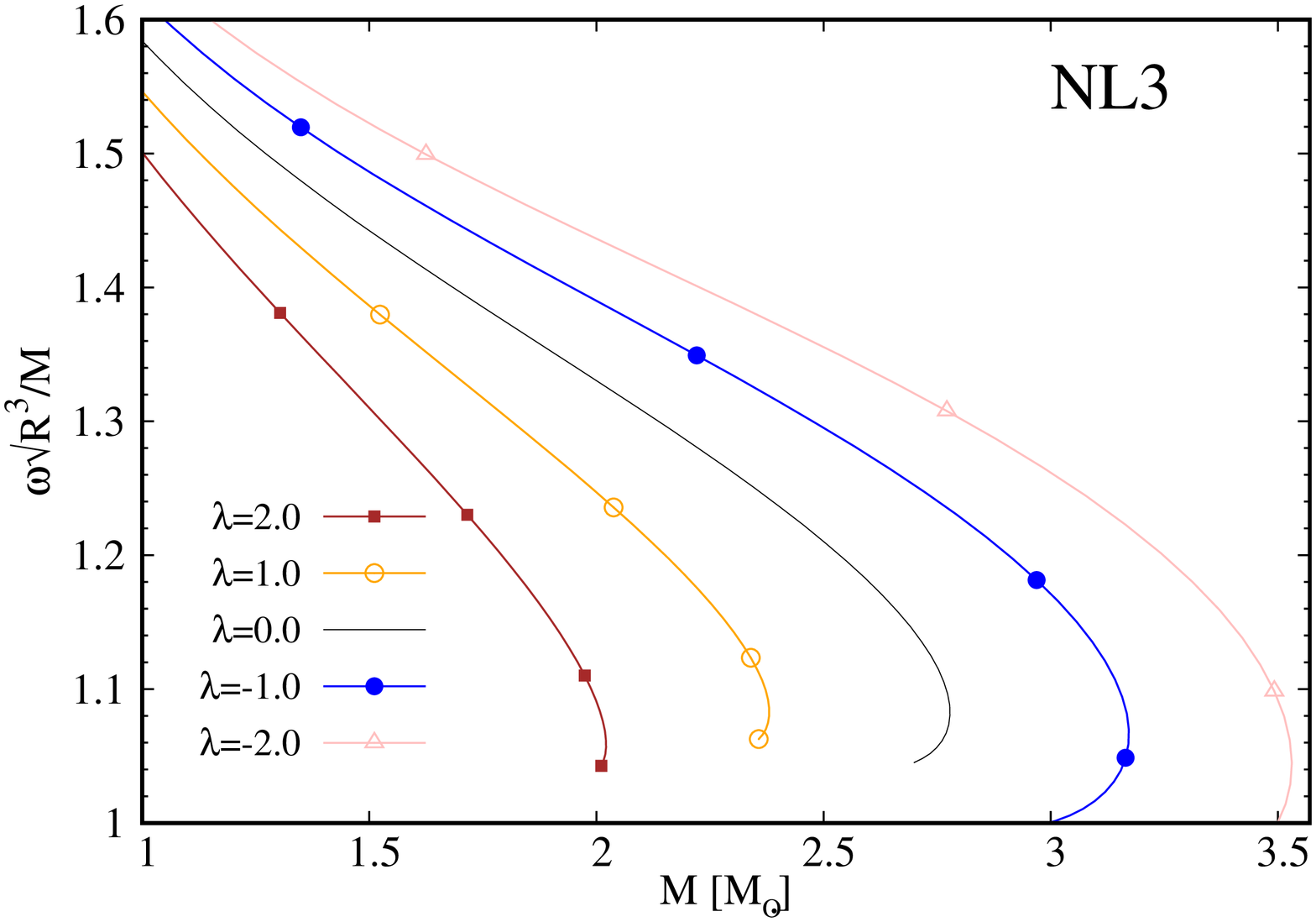}}
\caption{The normalized frequency $\omega$ as a function of the mass $M$ of anisotropic hadronic stars. The results are shown for several values of the parameter $\lambda$ using GM1 (left panel) and NL3 (right panel) parametrizations.}
\label{fig:h5}
\end{figure*}

In Fig. \ref{fig:h5} we show the normalized frequency $\omega\sqrt{R^{3}/M}$ as a function of the stellar mass, obtained for the GM1 (\ref{fig:h5a}) and NL3 (\ref{fig:h5b}) parameterizations. In both figures, we note that the normalized frequency can be higher (lower) when the parameter $\lambda$ is negative (positive). Consequently, for large values of $\vert\lambda\vert$ and large masses, the oscillation frequencies differ significantly from the case of the isotropic hadronic stars ($\lambda=0$). That result is observed for the GM1 and NL3 models.

\begin{figure*}[]
\subfloat[]{\label{fig:h6a}\includegraphics[angle=-90,width=0.47\textwidth]{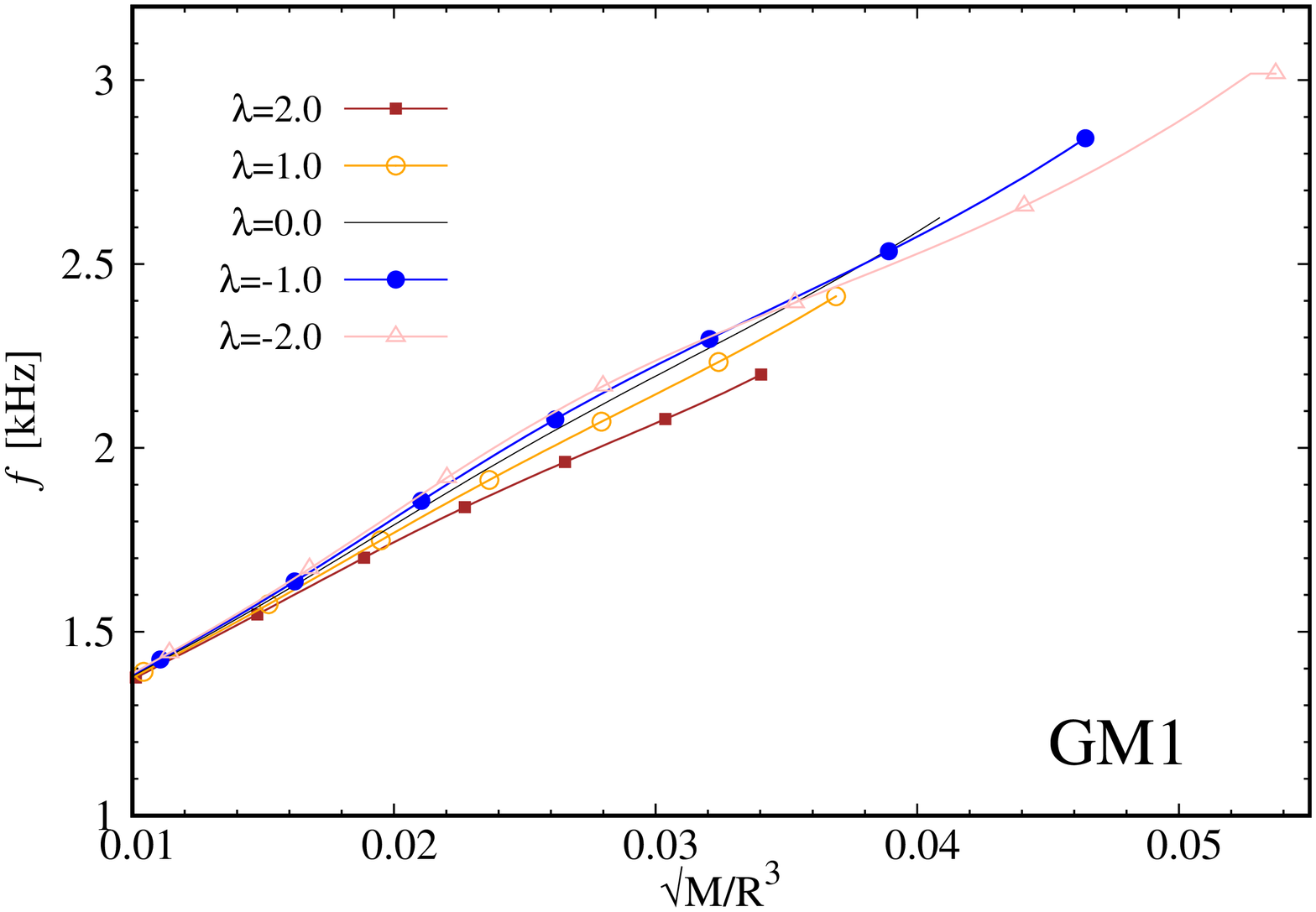}}\hspace{\fill}
\subfloat[]{\label{fig:h6b}\includegraphics[angle=-90,width=0.47\textwidth]{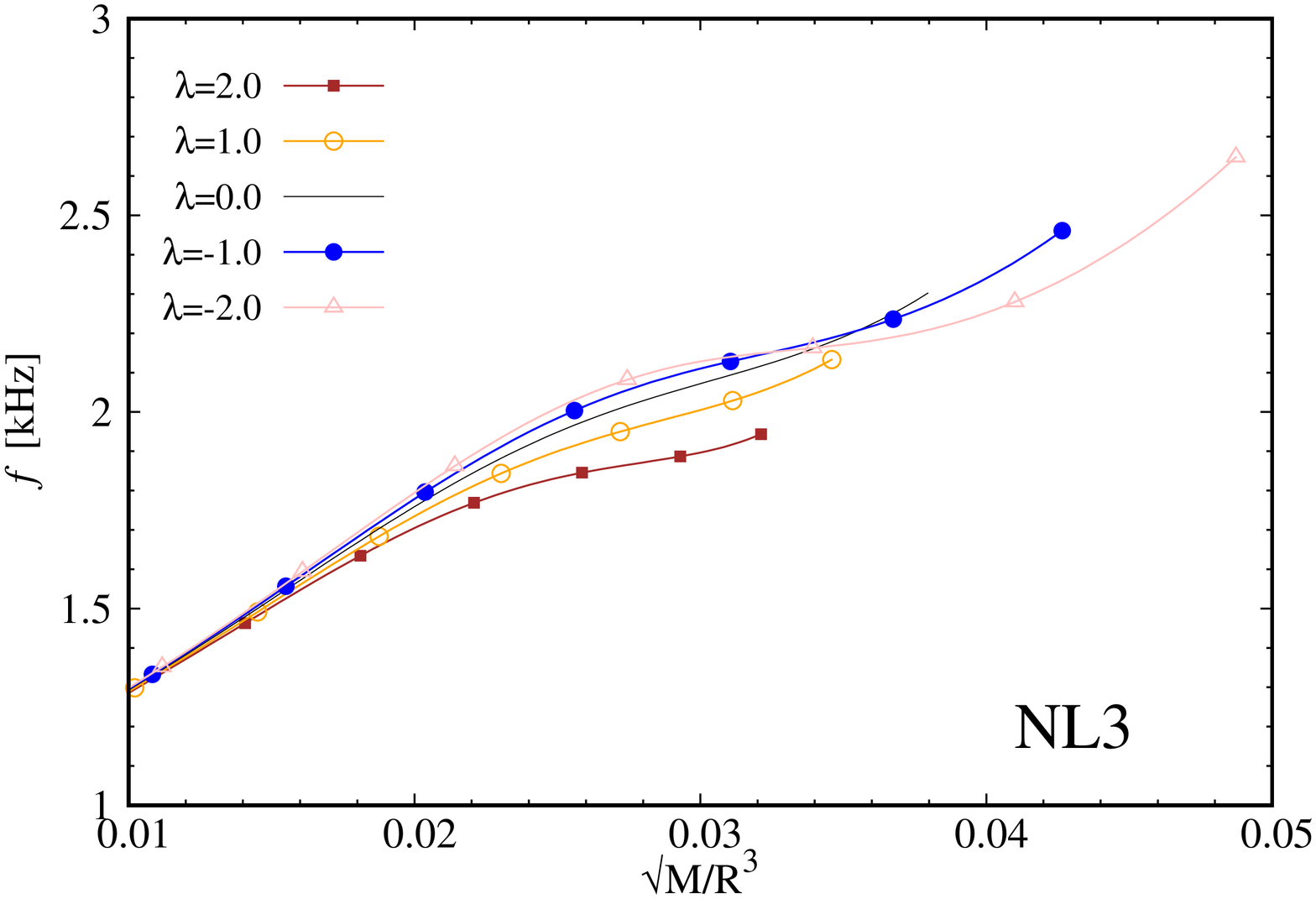}}
\caption{The fundamental mode frequency $f$  as a function of square root of the average density $\sqrt{M/R^3}$ for several values of the parameter $\lambda$. The results are shown for GM1 (left panel) and NL3 (right panel) parametrizations.}
\label{fig:h6}
\end{figure*}

Finally in Fig. \ref{fig:h6} we show the frequency of the fundamental mode as a function of the square root of the average density. The figure shows that the frequency does not change significantly for values lower than $0.02$\-~$\mathrm{km}^{-1}$ in the average density, i.e., in that region it is not possible to discriminate between isotropic and anisotropic hadronic stars by the use of the fundamental mode frequencies. Concerning the GM1 model (\ref{fig:h6a}), we realize that for average densities above $0.02$~$\mathrm{km}^{-1}$ and for large positive values of $\lambda$, the deviation of the anisotropic curves from the isotropic curves becomes more significant. Likewise, for the NL3 model (\ref{fig:h6b}), as the square root of the stellar average density grows for large positive values of $\lambda$, the deviation of the fundamental mode frequencies begins to be noticed. It is interesting to see that for negative values of $\lambda$, the fundamental mode frequency of anisotropic hadronic stars does not exhibit an important deviation with respect to the frequency of the isotropic stars, provided $0.01<\sqrt{M/R^3}<0.035$.

\subsection{Quark stars}

\begin{figure*}[]
\subfloat[]{\label{fig:q1a}\includegraphics[angle=-90,width=0.47\textwidth]{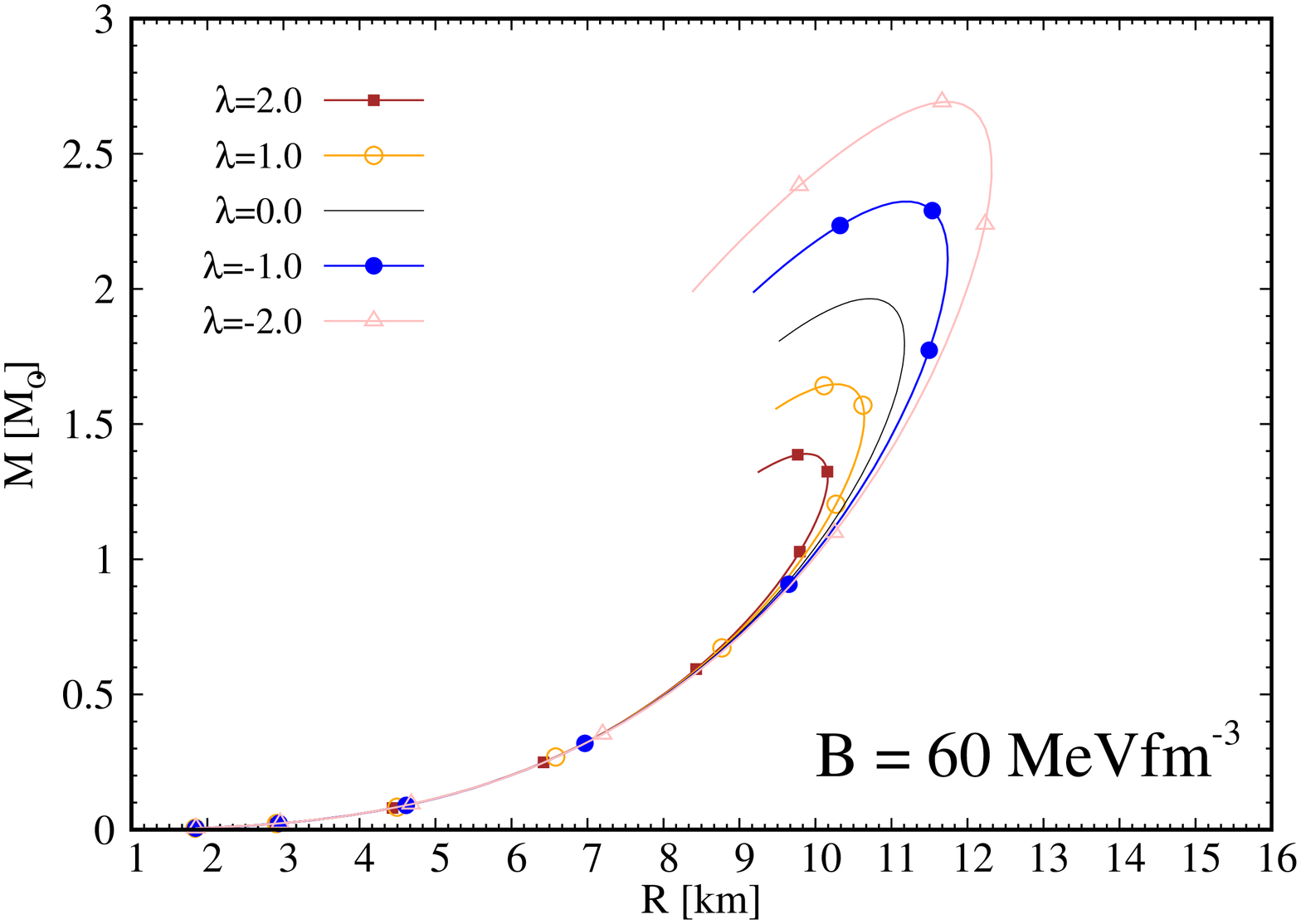}}\hspace{\fill}
\subfloat[]{\label{fig:q1b}\includegraphics[angle=-90,width=0.47\textwidth]{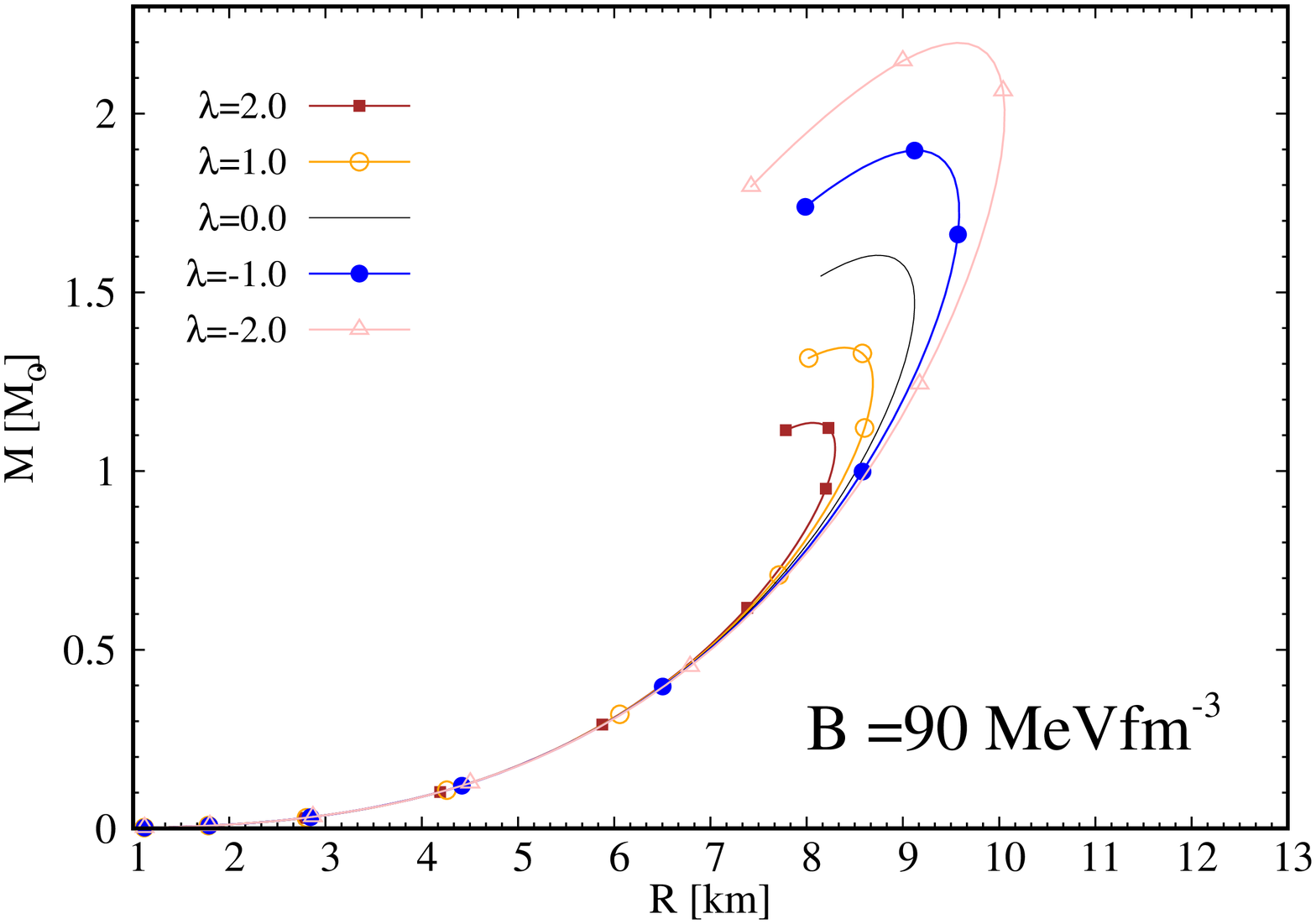}}
\caption{Mass-radius curves for anisotropic quark stars with Mit bag constant $B=60$\, $\mathrm{MeV}\mathrm{fm}^{-3}$ (left panel) and $90$\, $\mathrm{MeV}\mathrm{fm}^{-3}$ (right panel) for different values of the parameter $\lambda$.}
\label{fig:q1}
\end{figure*}

In Fig. \ref{fig:q1} we illustrate  the mass-radius relation for aniso\-tropic stars using the MIT bag model. In order to elaborate our results we employed two different values for the bag constant: $B=60$\, $\mathrm{MeV}\mathrm{fm}^{-3}$ (\ref{fig:q1a}) and $B=90$\, $\mathrm{MeV}\mathrm{fm}^{-3}$ (\ref{fig:q1b}). From that figure, we can appreciate how the equilibrium profile for strange stars is affected due to the anisotropy. For example, for the case $B=60$\, $\mathrm{MeV}\mathrm{fm}^{-3}$, the anisotropy has a much greater effect on masses above $1.0 M_{\odot}$.  We observe that when the parameter $\lambda$ changes, the values for the maximum mass change significantly. Similar results can be observed for the case of $B=90$\, $\mathrm{MeV}\mathrm{fm}^{-3}$.

\begin{figure*}[]
\subfloat[]{\label{fig:q2a}\includegraphics[angle=-90,width=0.47\textwidth]{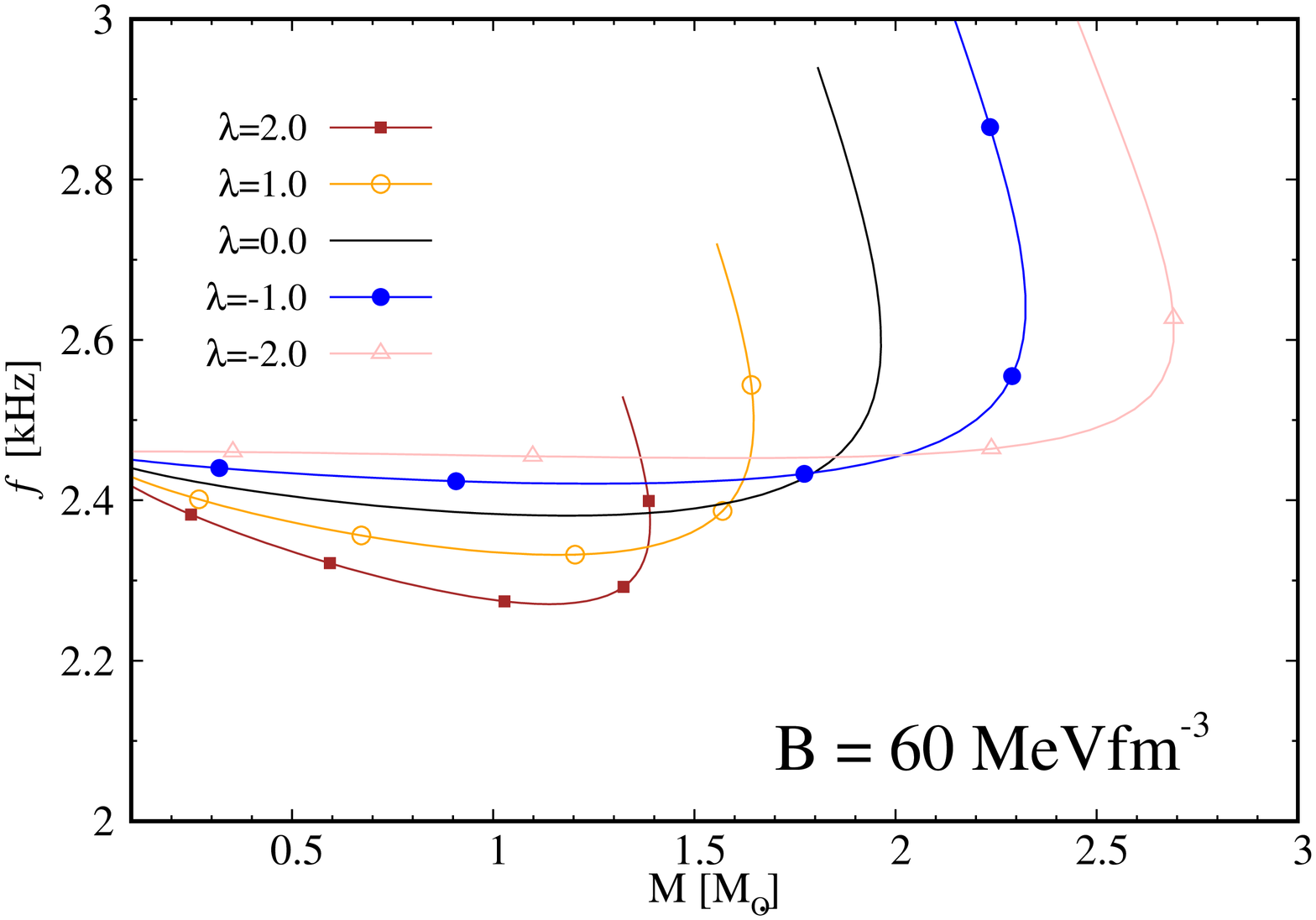}}\hspace{\fill}
\subfloat[]{\label{fig:q2b}\includegraphics[angle=-90,width=0.47\textwidth]{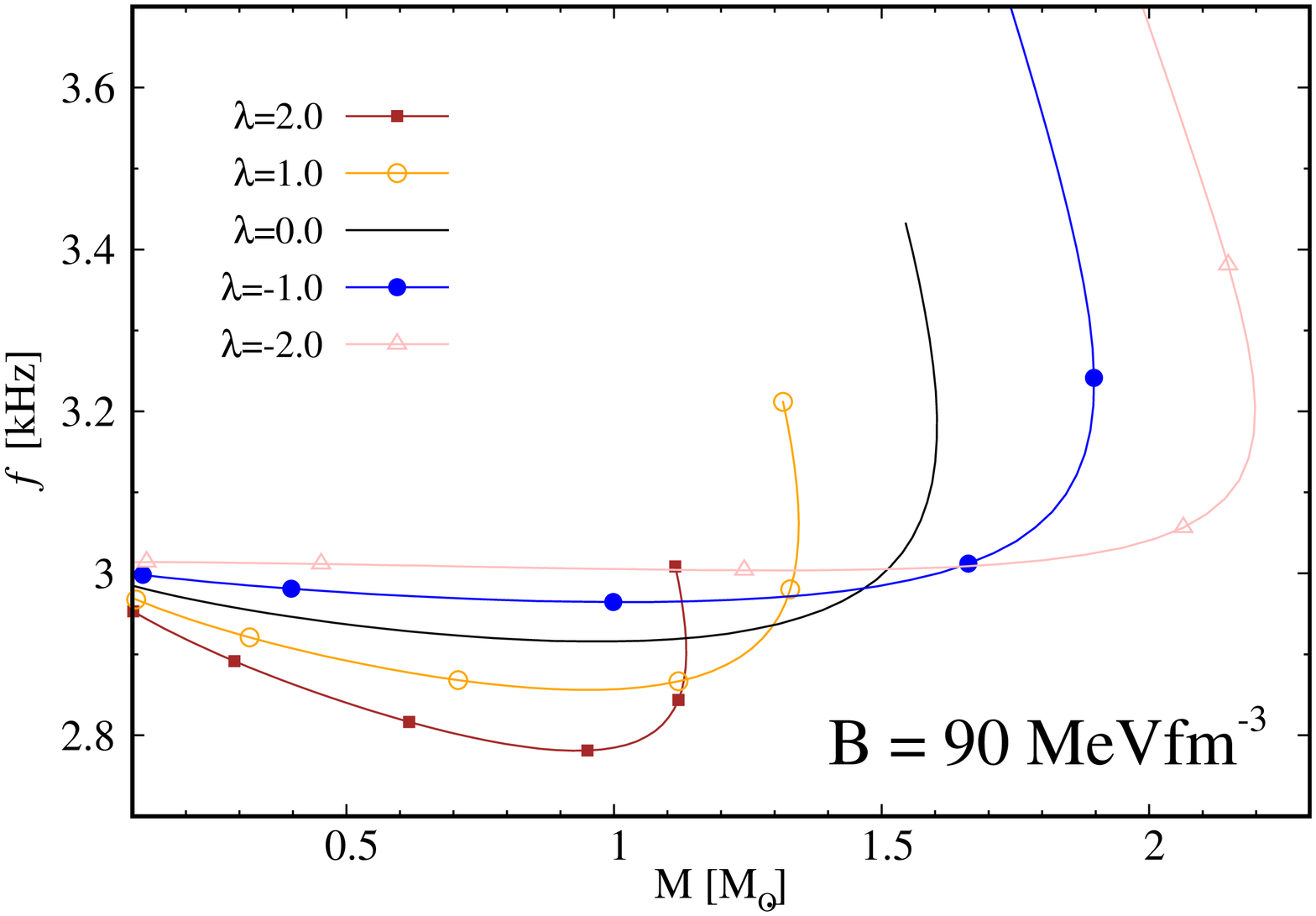}}
\caption{The fundamental mode frequency $f$ as a function of the mass $M$ with Mit bag constant $B=60$\, $\mathrm{MeV}\mathrm{fm}^{-3}$ (left panel) and $90$\, $\mathrm{MeV}\mathrm{fm}^{-3}$ (right panel). Results are plotted for different values of the parameter $\lambda$.}
\label{fig:q2}
\end{figure*}

In Fig. \ref{fig:q2} we show the profile of the frequency of the fundamental mode as a function of the stellar mass $M$. We observe that, when the absolute value of the $\lambda$ parameter changes, there is a considerable effect in aniso\-tropic stars when compared with the curves corresponding to isotropic (i.e. $\lambda =0$)  stars. Thus, the frequencies can be significantly higher or lower than in the case of isotropic strange stars, if $\lambda$ is slightly negative or positive.  For example, for a bag constant value of $B=60$\, $\mathrm{MeV}\mathrm{fm}^{-3}$ we see that the fundamental mode frequencies range from $2.3-3.0$~kHz.  For the case of positive values of the $\lambda $ parameter, anisotropic strange stars have low frequencies that decrease with the mass until some minimum value is reached. For the case of negative values of the $\lambda $ parameter, the frequencies behave nearly constant with the mass. Even though the anisotropic frequencies suffer big changes due to $\lambda$, the frequencies are lower than $4$~kHz \cite{MNRAS320:307:2001}.

\begin{figure*}[]
\subfloat[]{\label{fig:q3a}\includegraphics[angle=-90,width=0.47\textwidth]{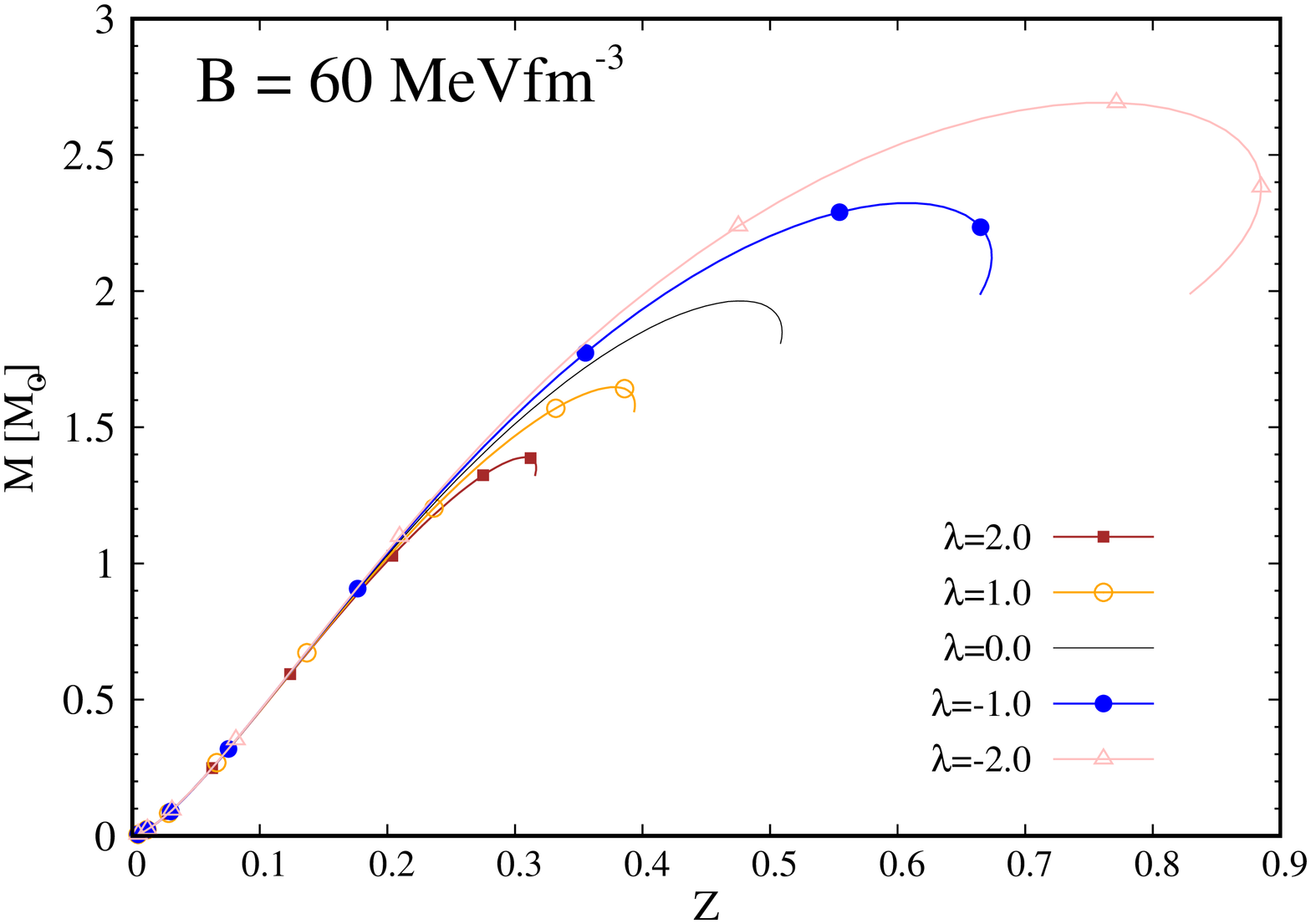}}\hspace{\fill}
\subfloat[]{\label{fig:q3b}\includegraphics[angle=-90,width=0.47\textwidth]{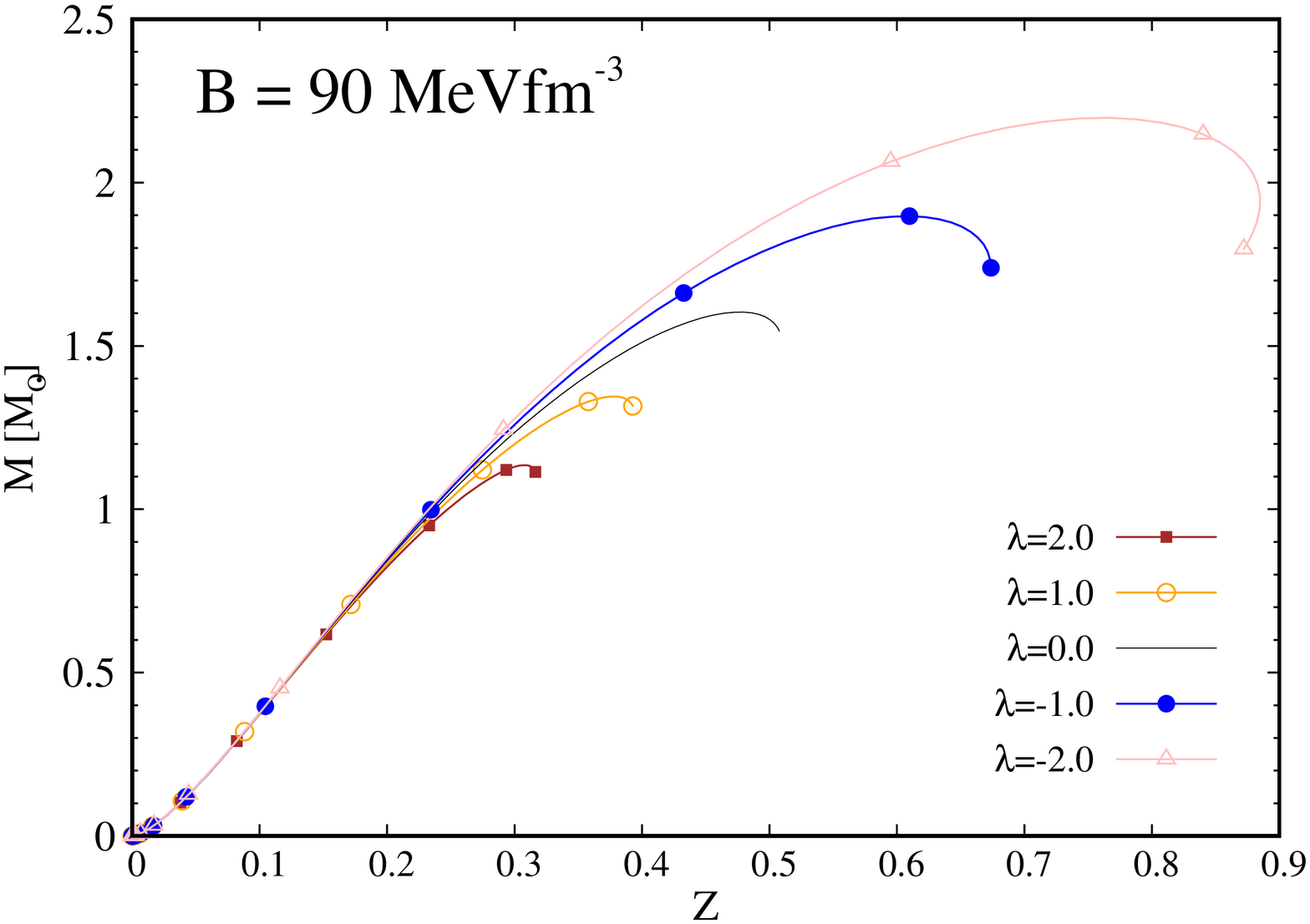}}
\caption{The total mass of anisotropic quark stars as a function of the redshift function $Z(R)=\mathrm{e}^{-\Phi(R)/2}-1$ for several values of the parameter $\lambda$. We use the Mit bag constant $B=60$\, $\mathrm{MeV}\mathrm{fm}^{-3}$ (left panel) and $90$\, $\mathrm{MeV}\mathrm{fm}^{-3}$ (right panel).}
\label{fig:q3}
\end{figure*}

Now we discuss the behavior of the gravitational redshift.  The results for the gravitational redshift as a function of the mass are showed in figure \ref{fig:q3}, for different values of $\lambda$. From the left panel, for $B=60$\, $\mathrm{MeV}\mathrm{fm}^{-3}$ , in Fig. (\ref{fig:q3a}), we observe that the gravitational redshift increases linearly for masses smaller than $\approx 1.4~M_{\odot}$, this means that for low mass configurations, the values of the the gravitational redshift becomes independent of the parameter $\lambda$. For stars with masses greater than $1.4~M_{\odot}$, the anisotropy produces significant changes for any $\lambda$. From the right panel, $B=90$\, $\mathrm{MeV}\mathrm{fm}^{-3}$, in Fig. (\ref{fig:q3b}), we observe that the gravitational redshift increases linearly for masses smaller than $\approx 1.1~M_{\odot}$, this means that for low mass configurations, the values of the the gravitational redshift becomes independent of the parameter $\lambda$. For stars with masses greater than $1.1~M_{\odot}$, the anisotropy produces significant changes for any $\lambda$. It is also clear that lower negative values of $\lambda$ give greater gravitational redshifts.

\begin{figure*}[]
\subfloat[]{\label{fig:q4a}\includegraphics[angle=-90,width=0.47\textwidth]{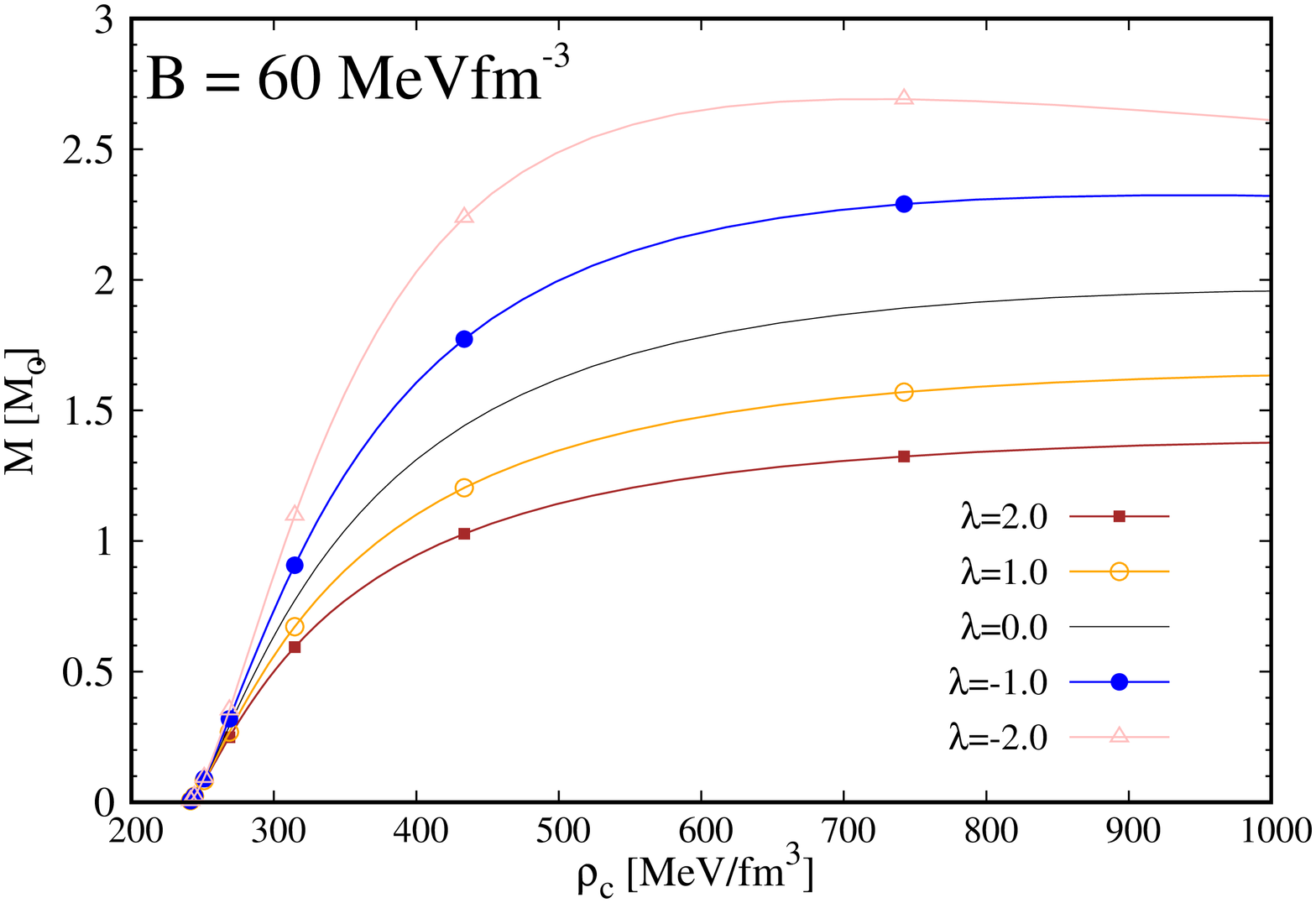}}\hspace{\fill}
\subfloat[]{\label{fig:q4b}\includegraphics[angle=-90,width=0.47\textwidth]{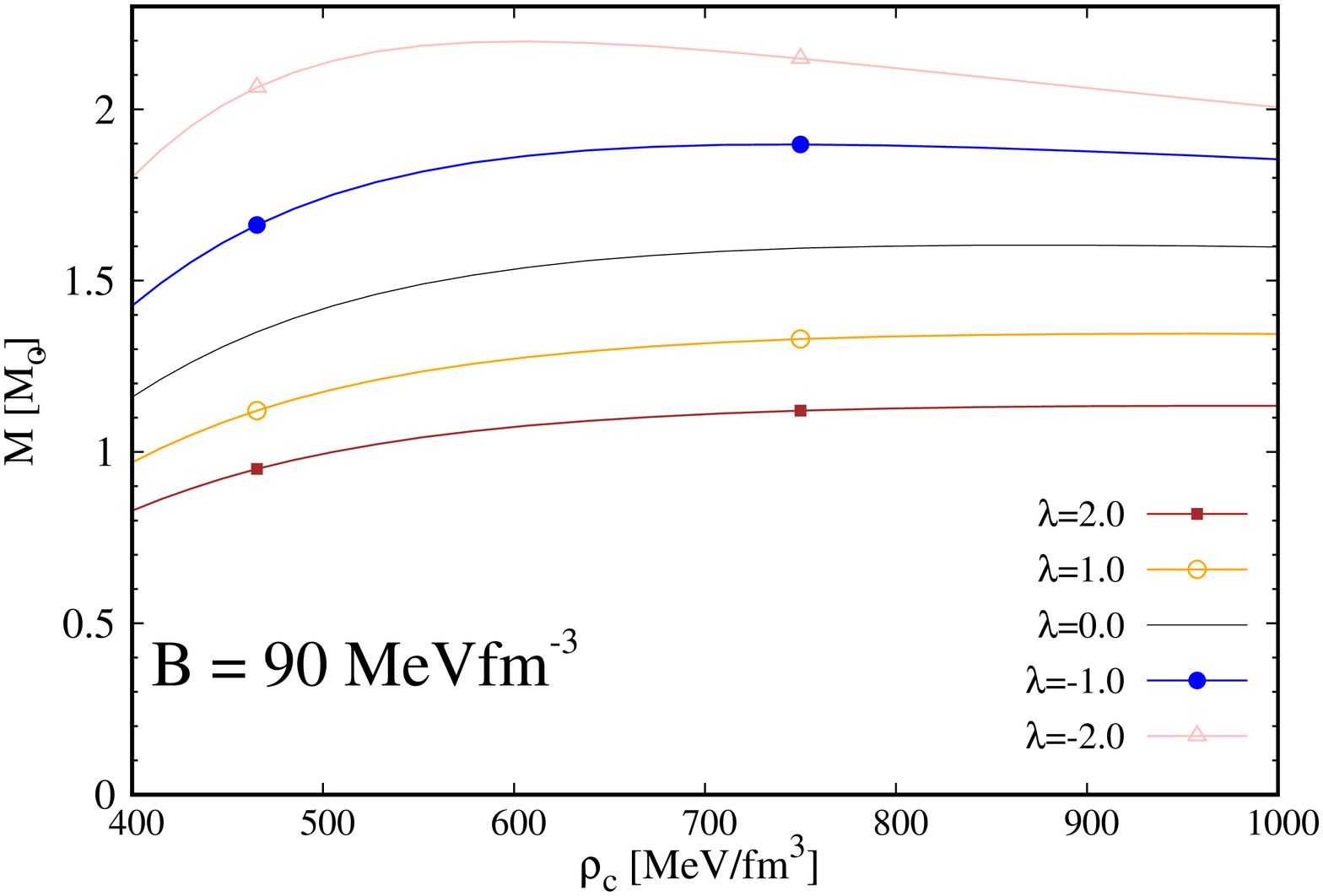}}
\caption{The total mass of anisotropic quark stars versus the central density $\rho_{c}$ with Mit bag constant $B=60$\, $\mathrm{MeV}\mathrm{fm}^{-3}$ (left panel) and $90$\, $\mathrm{MeV}\mathrm{fm}^{-3}$ (right panel) for different values of the parameter $\lambda$.}
\label{fig:q4}
\end{figure*}

In Fig. \ref{fig:q4}, the mass $M$ of anisotropic quark stars is shown as a function of the central density $\rho_c$ for different values of the parameter $\lambda$. For example, for the constant bag $B=60$\, $\mathrm{MeV}\mathrm{fm}^{-3}$, the anisotropic quark mass does not change considerably when $\lambda$ is changed for central densities below $300$~$\mathrm{MeVfm}^{-3}$. The stellar mass changes significantly for central densities above $\approx 300$~$\mathrm{MeVfm}^{-3}$ for any $\lambda$. Similar results can be observed for $B=90$\, $\mathrm{MeV}\mathrm{fm}^{-3}$.

\begin{figure*}[]
\subfloat[]{\label{fig:q5a}\includegraphics[angle=-90,width=0.47\textwidth]{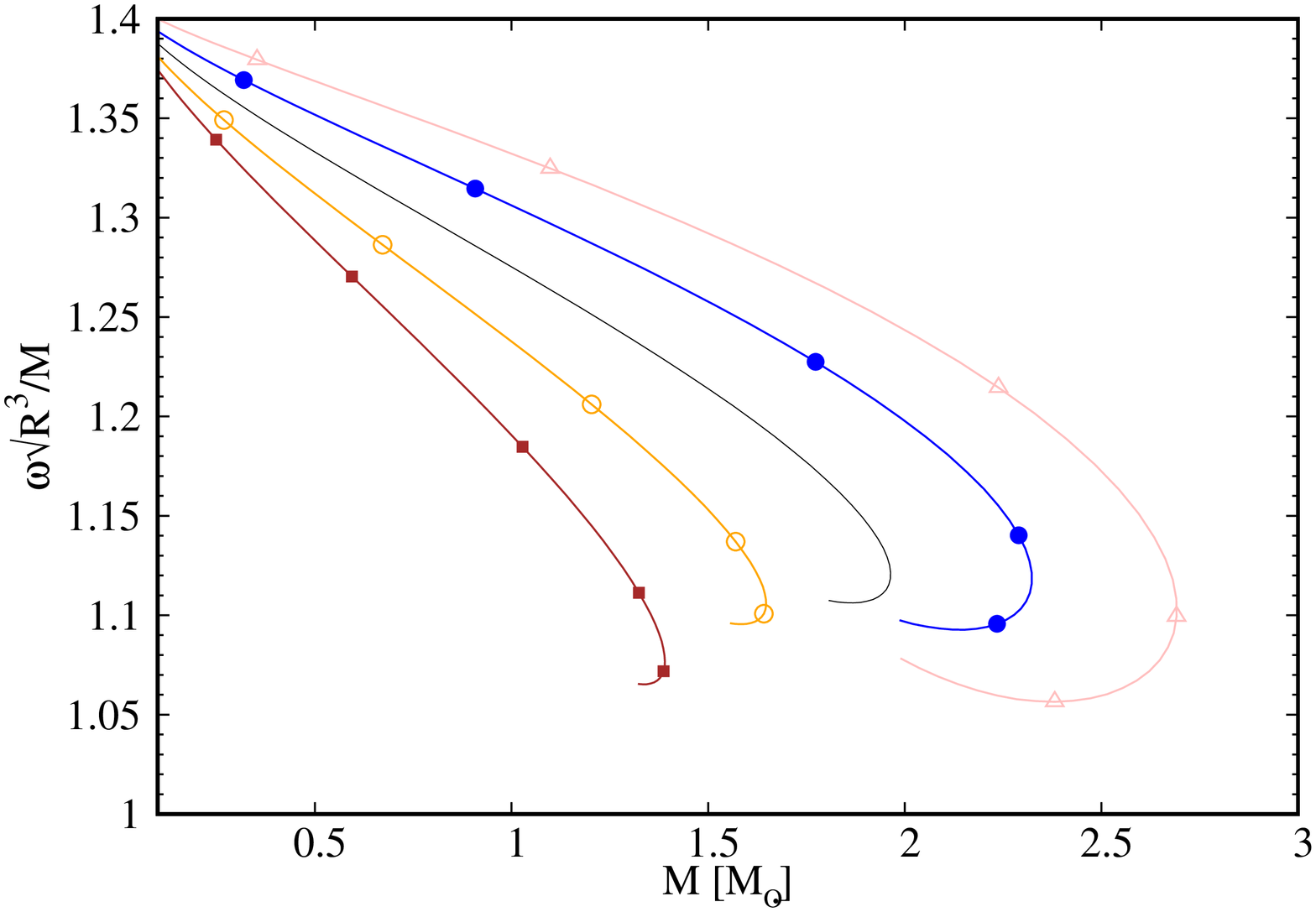}}\hspace{\fill}
\subfloat[]{\label{fig:q5b}\includegraphics[angle=-90,width=0.47\textwidth]{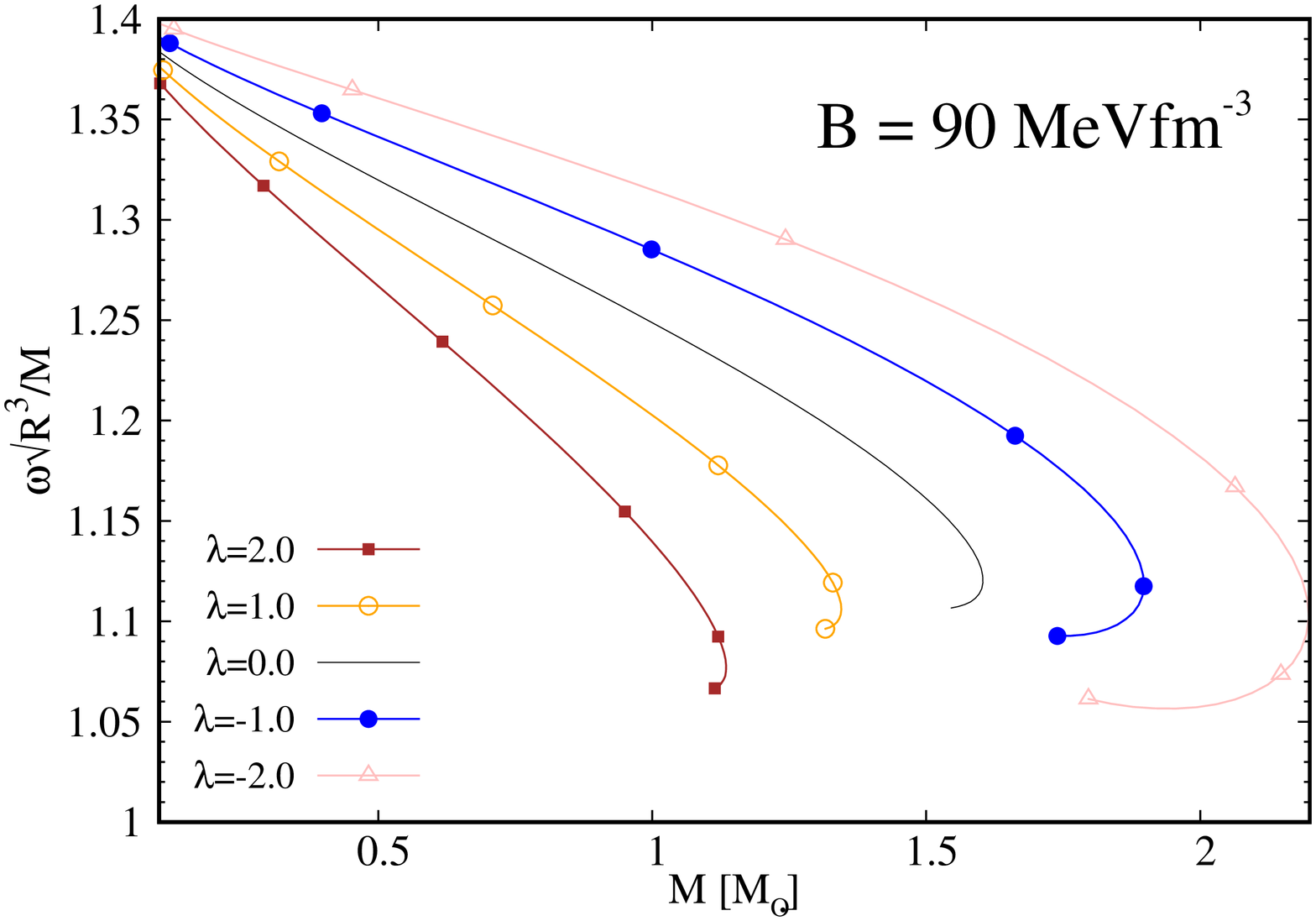}}
\caption{The normalized frequency $\omega$ as a function of the mass $M$ of anisotropic quark stars. The results are shown for several values of the parameter $\lambda$ with Mit bag constant $B=60$\, $\mathrm{MeV}\mathrm{fm}^{-3}$ (left panel) and $90$\, $\mathrm{MeV}\mathrm{fm}^{-3}$ (right panel).}
\label{fig:q5}
\end{figure*}

In Fig. \ref{fig:q5} we show the results for the normalized frequency of the fundamental mode $\omega\sqrt{R^3/M}$ as a function of the stellar mass . The analysis in this case is similar to the case considered in the Fig. \ref{fig:h5} (hadronic case), i.e., the frequencies change considerably when we increase the absolute value of $\lambda$. Also, we can say that for a fixed value of $\lambda$ and considering larger masses, any anisotropic curve is very different from the isotropic case. Consequently, the oscillation frequencies for massive stars can deviate considerably from the case of the isotropic strange stars whenever $\vert\lambda\vert$ takes large values. On the other hand, by comparing the frequency and the normalized frequency, we can notice that the maximum frequency value is $\approx 3.6$~kHz, while the maximum normalized frequency $\omega\approx 1.4$~kHz. Moreover, both the fundamental mode frequency and the normalized frequency decrease as the total mass increases for any value of $\lambda$. Later, they reach a minimum point and the star becomes unstable. While the decrease of anisotropic frequencies are a bit inclined, the anisotropic normalized frequencies are abruptly inclined for any $\lambda$. Another interesting observation is that, here, the curves don't overlap as it is shown for hadronic stars in the fi\-gu\-re \ref{fig:h5}.

\begin{figure*}[]
\subfloat[]{\label{fig:q6a}\includegraphics[angle=-90,width=0.47\textwidth]{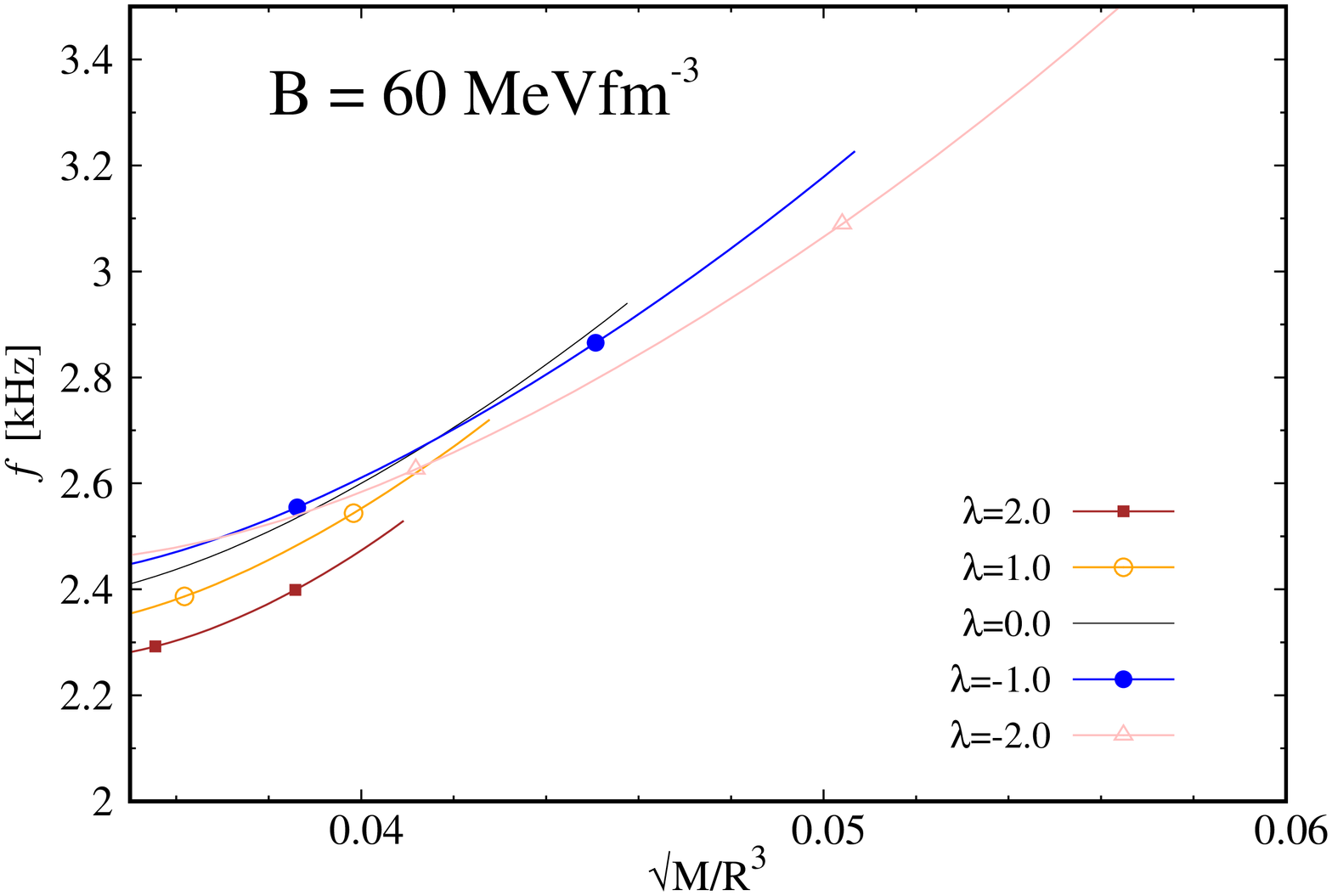}}\hspace{\fill}
\subfloat[]{\label{fig:q6b}\includegraphics[angle=-90,width=0.47\textwidth]{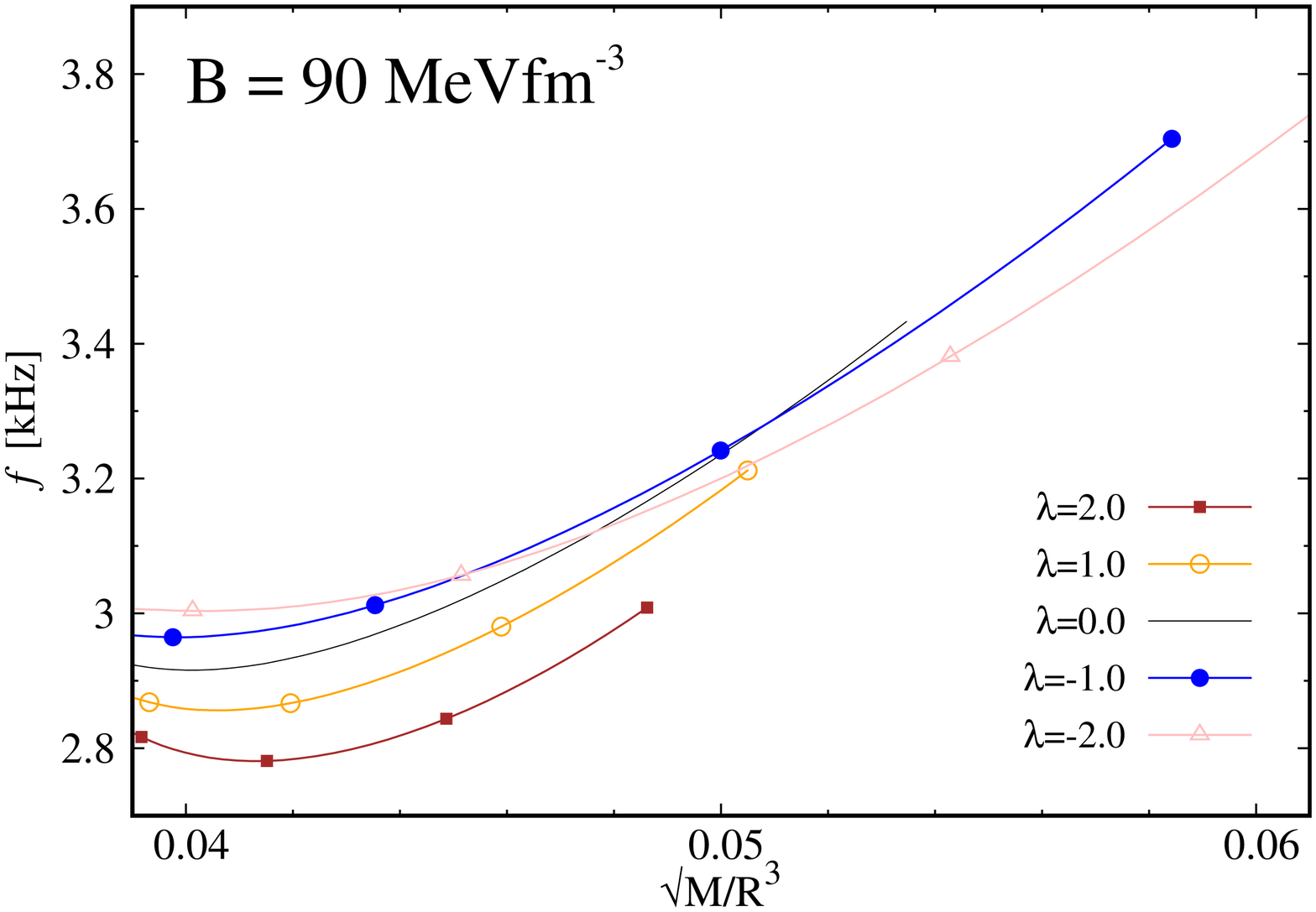}}

\caption{The fundamental mode frequency $f$ as a function of square root of the average density $\sqrt{M/R^3}$ for several values of the parameter $\lambda$.  The results are shown for Mit bag constant $B=60$\, $\mathrm{MeV}\mathrm{fm}^{-3}$ (left panel) and $90$\, $\mathrm{MeV}\mathrm{fm}^{-3}$ (right panel).}
\label{fig:q6}
\end{figure*}

The oscillation frequency of the fundamental mode as a function of the square root of the average density is shown in Fig. \ref{fig:q6}, for some values of the parameter $\lambda$. On the left panel (\ref{fig:q6a}), the plot shows small (big) changes of the frequencies from the isotropic (i.e., $\lambda =0$) strange stars only when $\lambda$ takes negative (positive) values and for small average density. However, the effect of anisotropy on the frequency begins to be noticed for any large value of the average density and for any negative $\lambda$. On the right panel (\ref{fig:q6b}), The situation of the deviation for the fundamental mode frequencies from the isotropic frequency curve are visualized better when the bag constant is increased, for any value of the parameter $\lambda$ and for any range along the average density. 

Our main results for each one of the EoS used in this work are summarized in Table \ref{TL2}.

\begin{table*}[ht]
\begin{center}
\caption{Neutron stars main properties for each one of the EoS used in the article.}
\label{TL2}
\begin{tabular}{ c c c c c c c c }
\hline
EoS & $\lambda$   &  $M$ ($M_{\odot}$) & R $(\mathrm{km})$  &  $f(\mathrm{kHz})$  & $Z$ & $\rho_c (\mathrm{MeV}\mathrm{fm}^{-3}) $  
& $\sqrt{M/R^3}$  (G)\\
\hline
MITB60& 2 & 1.38   & 9.88 & 2.37   & 0.3079 &  1316.929  & 0.037\\
MITB60&0 & 1.96   & 10.65 & 2.61  &  0.4814 &1219.027  & 0.043 \\
MITB60&-2  & 2.69 & 11.66 & 2.62 &  0.7715 & 742.395 & 0.041 \\
\hline
MITB90& 2 &  1.13 &  8.03&  2.91&  0.3096& 1022.144 & 0.046\\
MITB90& 0 & 1.60 & 8.71 & 3.19   &  0.4795  & 897.558& 0.049 \\
MITB90& -2 &  2.19& 9.56 &  3.20 & 0.7647  &  666.990&  0.051 \\
\hline
GM1& 2 & 1.67 & 11.46& 2.17 & 0.3256 &  1497.773  &  0.033 \\
GM1& 0 & 2.30 & 11.80 & 2.47  & 0.5360 &  1187.773  & 0.037   \\
GM1& -2 & 2.98 & 12.78 & 2.46  & 0.7973&   668.297 &  0.037 \\
\hline
NL3& 2 & 2.02 & 12.78 & 1.91  &  0.3693 &  1094.099  & 0.031 \\
NL3& 0 & 2.77 &  13.16 &  2.18 & 0.6292 &  897.558  & 0.034 \\
NL3& -2 & 3.53 & 13.97  &  2.18 & 0.9844&  527.829  &  0.035  \\
\hline
\end{tabular}
\end{center}
\end{table*}

\subsection{Comparison between theory and observations}

\begin{figure*}[]
\subfloat[]{\label{fig:h1ca}\includegraphics[angle=0,width=0.52\textwidth]{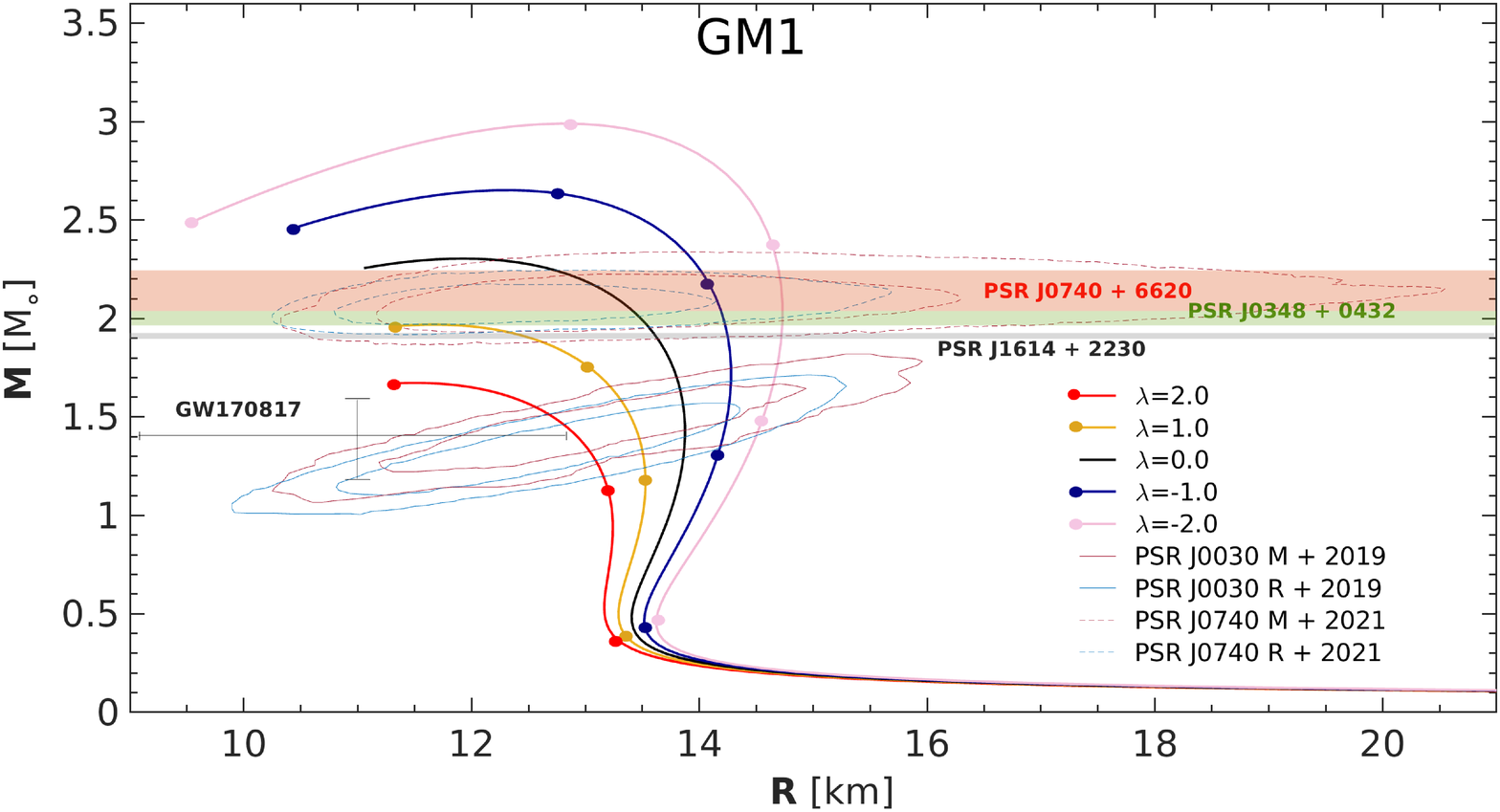}}\hspace{\fill}
\subfloat[]{\label{fig:h1cb}\includegraphics[angle=0,width=0.52\textwidth]{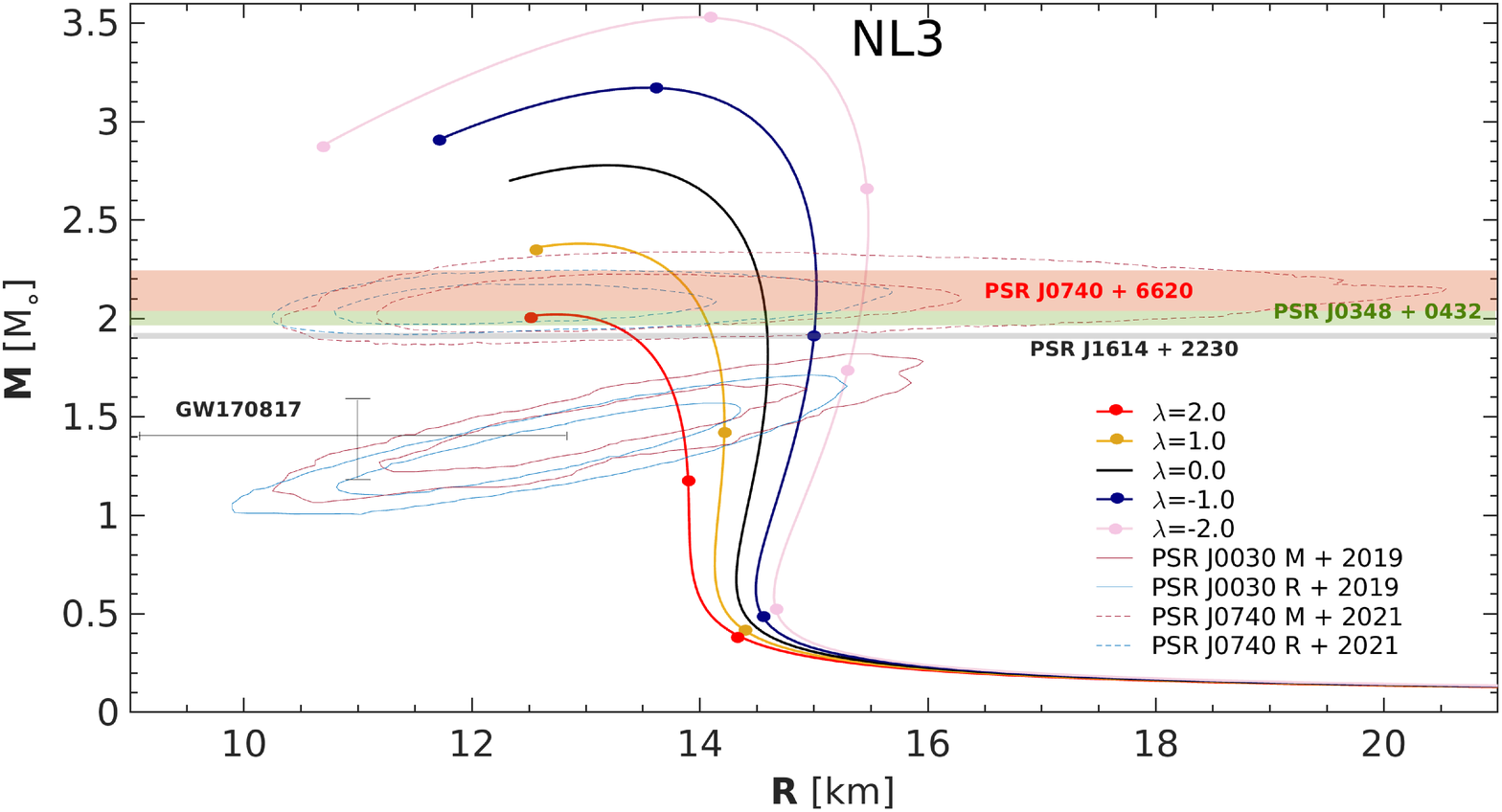}}
\caption{ Comparison of mass-radius curves for anisotropic hadronic stars with observational data using the GM1 (left panel) and NL3 (right panel) parametrizations for different values of the parameter $\lambda$. Black lines ($\lambda = 0$) correspond to isotropic hadronic stars.}
\label{fig:h1c}
\end{figure*}

\begin{figure*}[]
\subfloat[]{\label{fig:q1ca}\includegraphics[angle=0,width=0.52\textwidth]{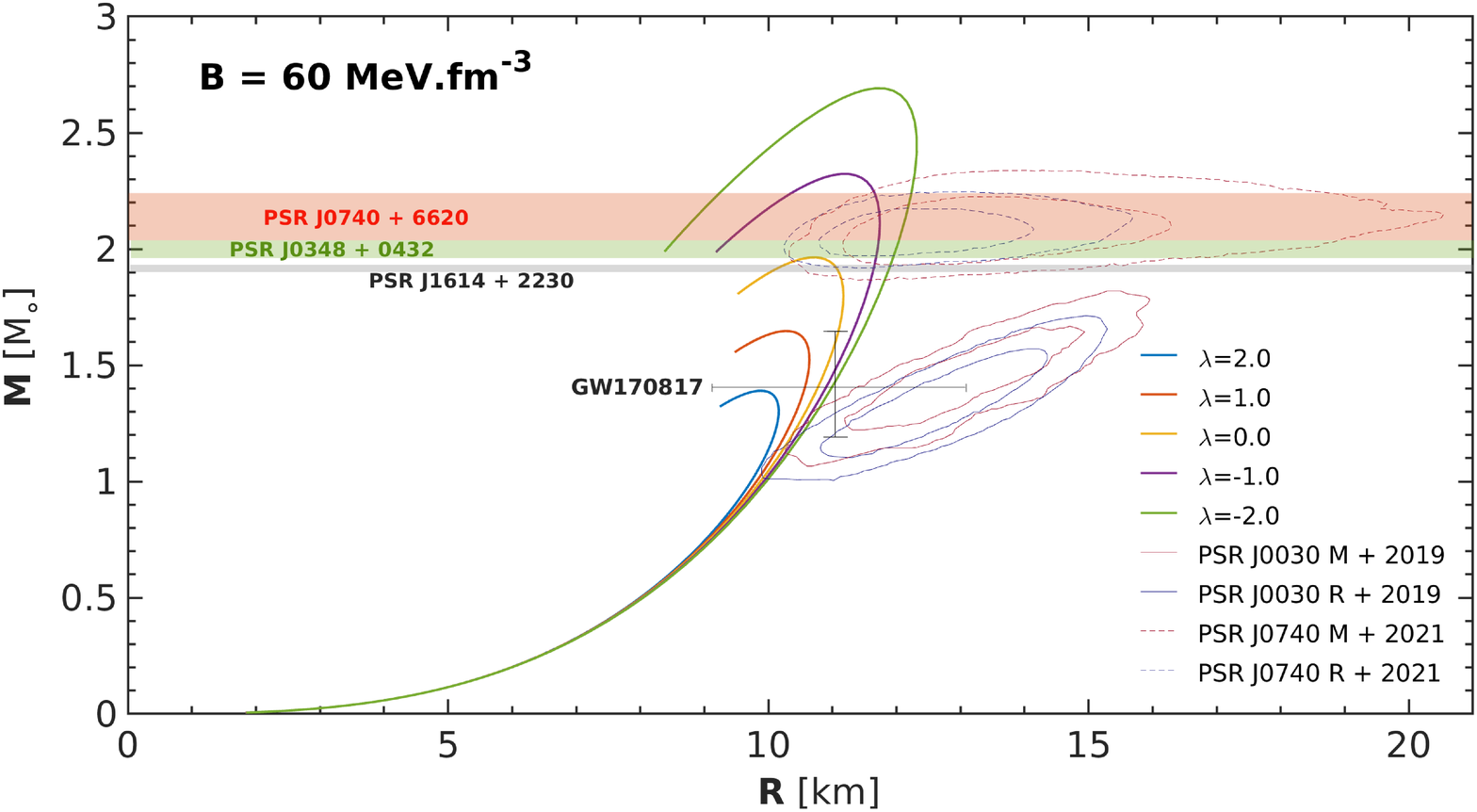}}\hspace{\fill}
\subfloat[]{\label{fig:q1cb}\includegraphics[angle=0,width=0.52\textwidth]{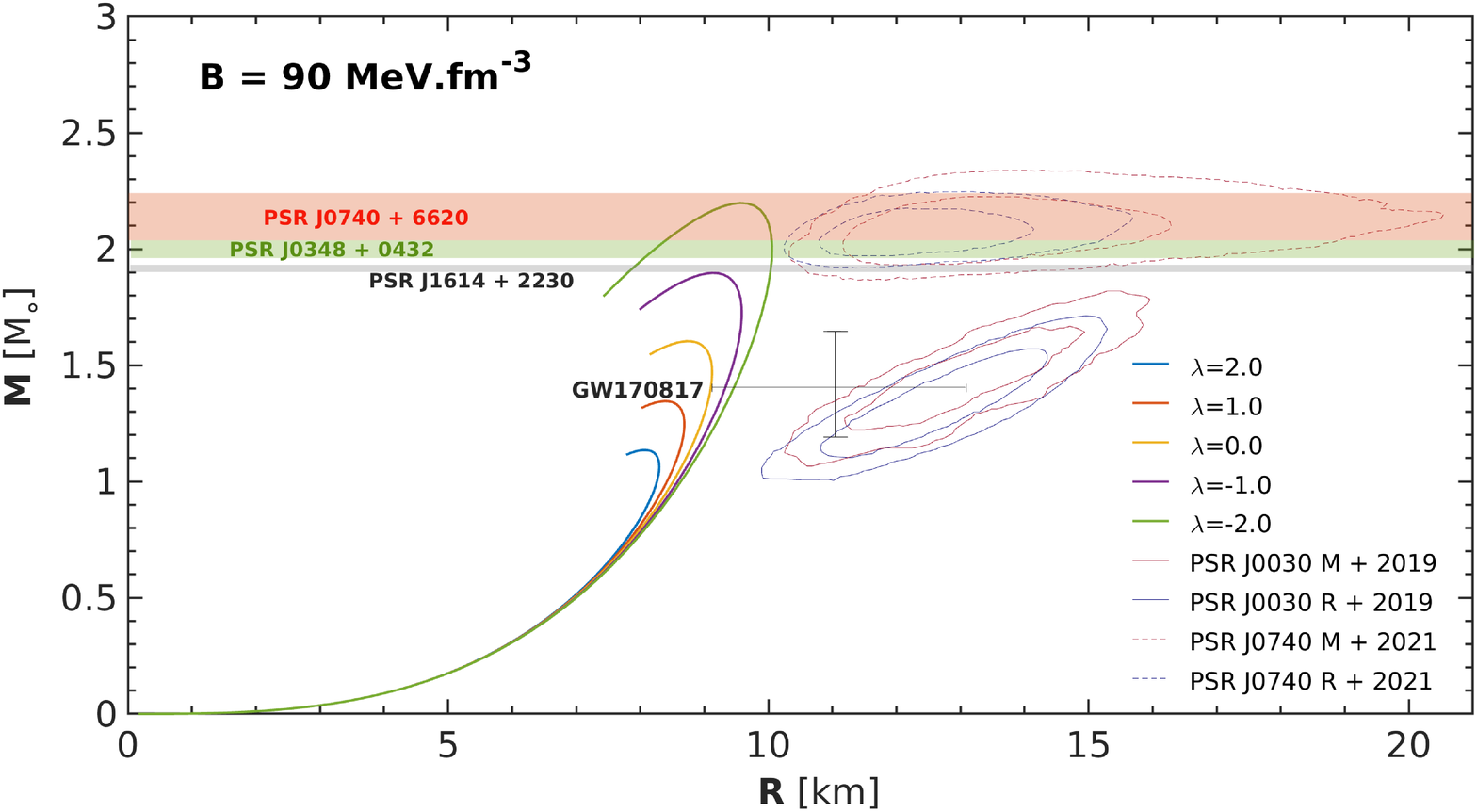}}
\caption{Comparison of mass-radius curves for anisotropic quark stars with observational data using Mit bag constant $B=60$\, $\mathrm{MeV}\mathrm{fm}^{-3}$ (left panel) and $90$\, $\mathrm{MeV}\mathrm{fm}^{-3}$ (right panel) for different values of the parameter $\lambda$.}
\label{fig:q1c}
\end{figure*}

Comparing the theory and its predictions with the observational data is an interesting test that validates whether or not an EoS is adequate to describe a realistic neutron star. In figures (\ref{fig:h1c}) and (\ref{fig:q1c}), we compared our results with the NICER constraints obtained from the pulsars PSRJ$0030+0451$ (Riley et al. 2019 \cite{Riley2019}; Miller et al. 2019 \cite{Miller2019}) and PSR J$0740+6620$ (Riley et al. 2021 \cite{Riley2021}; Miller et al. 2021 \cite{Miller2021}) for hadronic and quark stars, respectively. We also show the corresponding bands of the pulsars PSRJ$0740+6620$ \cite{Cromartie2020}, PSRJ$0348+0432$ \cite{S340:2013} and PSRJ$1614+2230$ \cite{Demorest2010} and an error bar corresponding to the binary system GW$170817$ \cite{PRL121:161101:2018}. For hadronic stars within the NL3 model, Fig. (\ref{fig:h1cb}), the curve related to the anisotropic factor $\lambda=1.0$ is the unique that satisfies the constraints obtained from Pulsars observations. Additionally, we can note that the decrease in the anisotropic factor produces an increase in the radius for the same stellar mass, this behaviour pushes out the models from the more central regions. The same effect can be observed for the GM1 model, but in this case an approximately vanishing anisotropic factor can be fulfil the above mentioned constraints shown in Fig. (\ref{fig:h1ca}). A good perspective can be found out when we look at Fig. (\ref{fig:q1ca}), where we test MIT bag model ($B=60$\, $\mathrm{MeV}\mathrm{fm}^{-3}$). Clearly, when $\lambda$ decreases the MIT bag model will tend to correspond to all observables. One can see that some of our results related to hadronic stars are more accurate and closer to empirical evidence of the neutron stars and therefore these EoS are great candidates for describing a realistic neutron star. At this stage, an interesting aspect related to our results is the fact that the anisotropic parameter $\lambda$ controls the values of the maximum star masses and the corresponding radii. This lead us to conclude that parameter $\lambda$ can be used as a tuning parameter to reproduce observational data of neutron stars for the observable mass and radius.

\section{Summary and Final Remarks}
\label{CONCLU}

In this article we examined how the non-radial oscillations of hadronic and quark stars are affected by aniso\-tropic effects. The results for hadronic stars were elaborated using two different parameterizations which are special cases of a
relativistic mean field model and for the case of quark stars we employed the MIT bag model.

All our analysis was made in the framework of the relativistic Cowling approximation, which means that the metric perturbations do not have dynamics when the stellar fluid is perturbed. Then we integrated numerically the non-radial oscillation equations obtaining the frequency of the fundamental mode, which it is considered to be more important for astrophysical applications. We even obtained more information by combining the frequency value of the fundamental mode with other physical properties of the star.

As far as we are concerned on hadronic and quark stars, we found that the anisotropy alters the following stellar properties: the frequency of oscillation of the fundamental mode, the total mass, total radius, the square root of the average density, the central density, the normalized frequency, and the surface redshift. Depending on the type of star to be considered, the influence of anisotropy is more obvious in some cases than in others, for instance, the frequency of the fundamental mode for hadronic stars has deviated slightly for small average density. The same small effect is observed to the gravitational redshift for both strange and hadronic masses below $1.5~M_{\odot}$; that is, it is not possible to distinguish between hadronic and quark stars. The same situation has happened for small quark masses as a relationship of the total radius. However, a big impact of anisotropy is clearly visualized on other relationships, such as, the frequency of the fundamental mode as function of the total mass has modified significantly the quark stars; and the same result is viewed for hadronic stars within the GM1 model. Another major impact exists for the normalized frequency as a function of mass as a result of anisotropy. Additionally, we have found noticeable effects of the anisotropy when we illustrate the hadronic total mass for radius lesser $14$~km within the GM1 model and a radius lesser $15$~km under the NL3 model. A similar situation is perceived for the total mass as a function of central density. One interesting feature that allows us to differentiate between strange stars and hadronic stars is that the former stars emit higher gravitational waves than the hadronic stars, i.e., anisotropy boosts higher values for the quark frequency of the frequency of the fundamental mode than hadronic frequency. In conclusion, all features mentioned lines above suggest that the anisotropic parameter $\lambda$ monitors all stellar properties of compact stars and it can be used as a tuning parameter to reproduce mass and radius observational data. From the comparison of our results with observational data, one can see that some of our theoretical results are in agreement with the empirical evidence of neutron stars, i.e. some EoS used in this work are great candidates for describing a realistic neutron star. In spite of some of our results related to hadronic stars are more accurate and closer to empirical evidence of the neutron stars, it is possible to realize the results with the MIT bag model can be favored depending on the bag constant $B$. The anisotropic factor $\lambda$ leads those curves to observable regions of the mass-radius diagram. Softer EoS, when not figured in experimental ranges, clearly can be favored when $\lambda$ decreases, which opens new possibilities for analysis of those EoS.

\begin{acknowledgements}
This work was partially supported by CNPq (Brazil) under grants 422755/2018-4 and 311925/2020-0, FAPEMA (Brazil) under grant UNIVERSAL-01220/18, and CAPES (Brazil). C. H. L. acknowledge Fundação de Amparo à Pesquisa do Estado de São Paulo (FAPESP) under Thematic Project No. 2017/ 05660-0 and Grant No. 2020/ 05238-9.
\end{acknowledgements}

\bibliographystyle{spphys}       

\end{document}